\documentclass[iopart12]{iopart}
\usepackage{amsfonts,amssymb,bm,youngtab}
\usepackage{amssymb}
\usepackage{epsfig}
\usepackage{subfigure}
\usepackage{graphicx}
\usepackage{color}
\usepackage{todonotes}
\usepackage{caption}
\usepackage{mathrsfs}
\usepackage{float}
\usepackage{hhline}
\usepackage{array}
\newcolumntype{C}{>{$}c <{$}}
\newcolumntype{L}{>{$}l <{$}}
\newcolumntype{R}{>{$}r <{$}}
\input{glyphtounicode.tex}
\def\beq{\begin{equation}}
\def\eeq{\end{equation}}
\def\beqa{\begin{eqnarray}}
\def\eeqa{\end{eqnarray}}
\def\hf{\textstyle{1\over2}}
\def\3hf{\textstyle{\frac{3}{2}}}
\def\half{{1\over2}}

\newcommand{\ket}[1]{\vert #1 \rangle}
\newcommand{\cg}[6]{\mbox{$\Bigl\langle {#1 \atop #2} ; {#3 \atop #4} \,\big\vert\, {#5 \atop #6} \Bigr\rangle$}}
\newcommand{\rcg}[6]{\mbox{$\Bigl\langle{#1 \atop #2} ;{ #3 \atop #4} \, \Vert\, {#5 \atop #6}\Bigr\rangle $}}
\newcommand{\sixj}[6]{\mbox{$\left\{\begin{array}{ccc} #1 & #2 & #3\\ #4 & #5 & #6 \end{array}\right\}$}}
\newcommand{\bra}[1]{\langle #1 \vert}
\newcommand\bigket[2]{\left\vert #1\atop #2\right\rangle}
\newcommand\bigbra[4]{\left\langle {#1\atop #2}; {#3 \atop #4} \right\Vert}
\newcommand\unit{\mathinner{\hbox{1}\mkern-4mu\hbox{l}}}
\newcommand\rket[1]{\vert\, #1 )}
\renewcommand\rme[3]{\langle #1 \,\Vert\, #2 \,\Vert \, #3\rangle}

\newcommand{\binom}[2]{ {#1 \choose #2}}

\newcommand{\myfrac}[2]{\leavevmode\kern.1em\raise.5ex\hbox{\scriptsize
$#1$}\kern-.1em {\scriptsize
/}\kern-0.10em\lower.25ex\hbox{\scriptsize $#2$}}
\newcommand{\mysmallfrac}[2]{\leavevmode\kern.1em\raise.5ex\hbox{\tiny
$#1$}\kern-.1em {\tiny
/}\kern-0.10em\lower.25ex\hbox{\tiny$#2$}}

\newcommand{\hdg}[1]{\textcolor{black}{#1}}

\def\sugroup{\mathfrak{su}}
\newcommand{\tab}{\hspace{10mm}}

\begin{document}

\title{$SU(3)$ Clebsch-Gordan coefficients and some of their symmetries}
\author{Alex Cl\'{e}sio Nunes Martins, Mark W. Suffak, Hubert~de~Guise}

\address{ Department of Physics, Lakehead University, Thunder
Bay, Ontario P7B 5E1, Canada}

\begin{center}
\today
\end{center}

\begin{abstract}
We discuss the construction and symmetries of $\sugroup(3)$ Clebsch-Gordan coefficients arising from $\sugroup(3)$ basis states constructed as triple tensor products
of two-dimensional harmonic oscillator states.  Because of the $\sugroup(2)$ symmetry of the basis states, matrix elements and recursion relations are easily expressed in terms of $\sugroup(2)$ technology.  As the Weyl group has a particularly simple action on these states, Weyl symmetries of the $\sugroup(3)$ coupling coefficients generalizing the well known $m\to -m$ symmetry of $\sugroup(2)$ coupling can be obtained, so that any coefficient can be obtained as a sum of Weyl-reflected coefficients lying in the dominant Weyl sector.  Some important cases of multiplicity-free \hdg{decompositions} are discussed as examples of applications.
\end{abstract}

\section{Introduction and basic result}

Clebsch-Gordan (CG) coefficients occur regularly in physics in a variety of situations, from angular momentum coupling \cite{varshalovich1988quantum}\cite{rose1995elementary}\cite{edmonds2016angular} to \hdg{direct product of} 
representations of various other groups 
(see for instance \cite{king2002multiplicity}\cite{van1997coupling}\cite{caprio2009construction}\cite{rowe2004spherical}\cite{hecht1983some})
 including 
representations of the symmetric group \cite{chen2002group} and couplings in tensor networks \cite{weichselbaum2012non}.  
The symmetries of these coefficients is not only of practical interest for economy of resources \cite{everdingcalculation} but also because they 
imply some relations (sometimes unexpected) \cite{derome1966symmetry}
\cite{smorodinskii1972clebsch}\cite{smorodinskii1978clebsch} between states connected by these symmetries, as best exemplified by the 
so-called Regge symmetries for
$\sugroup(2)$ coupling coefficients\cite{regge1958symmetry}.

 In this paper we present an easy algorithm for the evaluation of $\sugroup(3)$ Clebsch-Gordan coefficients.  In the specific case where the 
 \hdg{decomposition of the direct product} is multiplicity free, i.e. when an irrep $(p_2,q_2)$ occurs
 at most once in the decomposition of $(p_1,q_1)\otimes (\lambda,\mu)$, we obtain semi-analytical expressions for $\sugroup(3)$ CGs, by which we mean that our expressions involve (reasonably) simple sums of terms familiar from $\sugroup(2)$ coupling theory.    Although not the most general case, multiplicity-free \hdg{decompositions} occurs whenever at least \emph{one} of $p_1,q_1,\lambda$ or $\mu$ is $0$; this situation therefore includes the useful case of repeated $N$-fold \hdg{direct products} such as $(p_1,0)^{\otimes N}$.

There already exists powerful algorithms for the numerical evaluation of $\sugroup(n)$ CG coefficients \cite{draayer1973wigner}\cite{alex2011numerical}\cite{bahri2004programs}.  Our results are of interest
because sums can be done symbolically so the final CGs are all exact square roots of rational numbers.  
This simplification of our final expression in Eq.~(\ref{eqn:generalexpression}) is the byproduct of working in a basis where 
$\sugroup(3)$ states are expressed
using unitary-unitary duality in terms of a triple \hdg{product} of angular momentum states, so that all
the pieces in our expressions are coefficients obtained from $\sugroup(2)$ coupling theory, all
of which are themselves square roots of rational expressions.  As a result, it is often possible to keep
some of the representation labels as parameters, and investigate various asymptotic limits \cite{stancu20163}\cite{de1995asymptotic}\cite{de1998clebsch} beyond the tables provided, for instance,
by Hecht \cite{hecht1965su3}.

Finally, because the action of the Weyl group on these states is easily expressible in terms of $\sugroup(2)$ $6j$-symbols,  we
dispense with the need to exponentiate some transformations \cite{everdingcalculation} to obtain the action
of the permutation group on basis states and easily obtain Weyl relations between $\sugroup{(3)}$ CGs.

\section{Basis states and their matrix elements}

\hdg{In this section we introduce the notation in terms of occupation numbers and the matrix elements of
generators and powers of generators.  This is essential because the Weyl group acts naturally on this ``occupation basis'' and we need matrix elements (or at least reduced matrix elements of appropriate tensors) to speed up the computation
of CG coefficients.}

\subsection{$\sugroup(3)$ generators and basis states}  

Let
\begin{equation}
\hat C_{ij}:= \hat a^\dagger_{i1}\hat a_{j1}+\hat a^\dagger_{i2}\hat a_{j2}+\hdg{\hat a^\dagger_{i3}\hat a_{j3}}
\end{equation}
and note that $[\hat C_{ij},\hat C_{k\ell}]=\delta_{jk}\hat C_{i\ell}- \delta_{i\ell}\hat C_{kj}$ close on the algebra $\mathfrak{u}(3)$.  
\hdg{We will be working in an occupation basis where the operators 
$\hat C_{ii}= \hat a^\dagger_{i1}\hat a_{i1}+\hat a^\dagger_{i2}\hat a_{i2}+\hat a^\dagger_{i3}\hat a_{i3}$ are diagonal.} The $\sugroup(3)$ 
subalgebra is \hdg{obtained from the nine $\hat C_{ij}$} by selecting \hdg{as} Cartan elements $\hat h_1=\hat C_{11}-\hat C_{22}$ and $\hat h_2=\hat C_{22}-\hat C_{33}$, \hdg{and keeping the} set of raising
operators $\hat C_{12}, \hat C_{13}, \hat C_{23}$ and lowering operators $\hat C_{21},\hat C_{31},\hat C_{32}$.    

\begin{figure}[h!]
\centering
\includegraphics[scale=0.3]{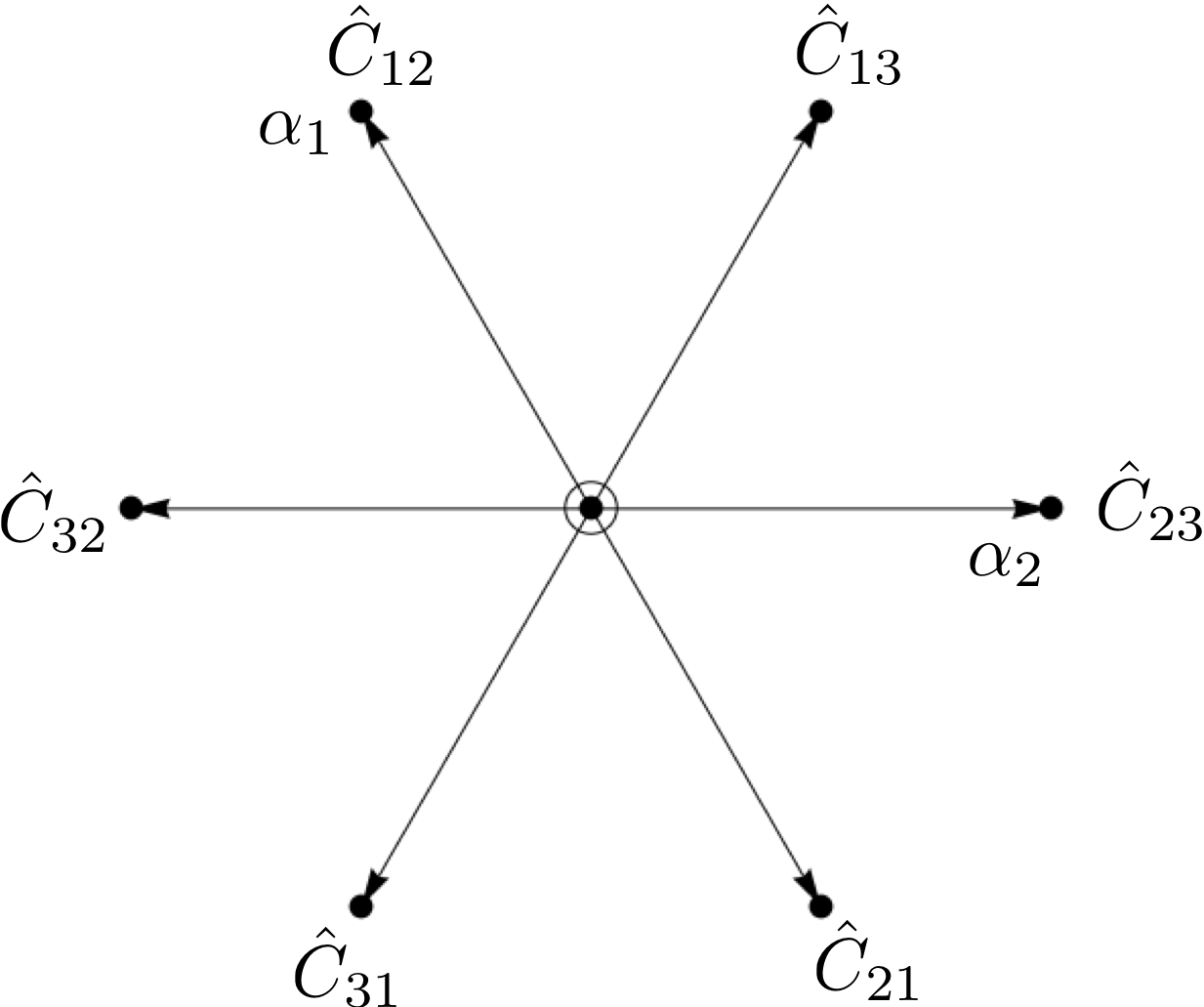}
\caption{The $\sugroup(3)$ root diagram showing the two \hdg{simple} roots, with the various generators with non-zero weight located on the root diagram.}
\label{fig:su3rootdiagram}
\end{figure}

The basis states for irrep $(\lambda,\mu)$ of $\sugroup(3)$ are constructed following \cite{rowe1999representations} as 
\begin{eqnarray}
&&\vert (\lambda,\mu)\nu; I\rangle \equiv \vert (\lambda,\mu)\nu_1\nu_2\nu_3;I\rangle \nonumber \\
&& := \sum_{\stackrel{m_1m_2}{m_3N}}
\cg{\frac{1}{2}\nu_3}{m_3}{\frac{1}{2}\nu_2}{m_2}{I}{N}
\cg{I}{N}{\frac{1}{2}\nu_1}{m_1}{\frac{1}{2}\lambda}{\frac{1}{2}\lambda}
\ket{\textstyle\frac{1}{2}\nu_1 m_1}\ket{\frac{1}{2}\nu_2 m_2}\ket{\frac{1}{2}\nu_3 m_3}
\label{eq:fullsu3states}
\end{eqnarray}
\hdg{where $\nu_1,\nu_2,\nu_3$ are non-negative occupation numbers subject to the constraint $\nu_1+\nu_2+\nu_3=\lambda+2\mu$.
These occupation numbers are eigenvalues of the number operators $\hat C_{ii}$. 
Each $\vert\frac{1}{2}\nu_i;m_i\rangle $ is itself an $\sugroup(2)$ state \cite{schwinger2015angular},} denoted by 
\begin{equation}
\vert \textstyle\frac{1}{2}\nu_i m_i\rangle = 
\displaystyle\frac{(\hat a_{i1}^\dagger)^{ \frac{1}{2}\nu_i+m_i}(\hat a_{i2}^\dagger)^{ \frac{1}{2}\nu_i-m_i}}
{\sqrt{( \frac{1}{2}\nu_i+m_i)!( \frac{1}{2}\nu_i-m_i)!}}\vert 0\rangle\, .
\end{equation}
\hdg{with $-\frac{1}{2}\nu_1\le m_i \le \frac{1}{2}\nu_1$ with $2m_i$ an integer.  The basis states for the irrep
$(\lambda,\mu)$ can be constructed without using $\hat a^\dagger_{j3}$, but $\hat{a}^\dagger_{j3}\hat a_{k3}$ are inserted in the 
definition of $\hat C_{jk}$ for the later convenience of some coefficient calculations.}  The coefficients 
$\cg{\frac{1}{2}\nu_3}{m_3}{\frac{1}{2}\nu_2}{m_2}{\hdg{I}}{N}$ {\it etc.} in Eq.(\ref{eq:fullsu3states}) are $\sugroup(2)$ CG coefficients.  
For notational convenience, 
$\nu$ will often be a shorthand to denote by definition the triple $(\nu_1\nu_2\nu_3)$; 
if some specific component of $\nu$ is needed, it will be explicitly indicated by the appropriate subscript.

The $\sugroup(3)$ weight of $\vert (\lambda,\mu)\nu;I \rangle$ is $(\nu_1-\nu_2,\nu_2-\nu_3)$.   
As constructed, the state also carries an irrep $\hdg{I}$ of the 
$\sugroup(2)\oplus \mathfrak{u}(1)$ subalgebra spanned by $\{\hat C_{23}, \hat C_{32}, \hat h_2,\hat h_1\}$.
The label $\hdg{I}$ is required to lift the ambiguity
arising when more than one state share the same weight.

The 
$\mathfrak{u}(1)$ generator is taken to be $\hat H_1=2\hat h_1+\hat h_2=2\hat C_{11}-\hat C_{22}-\hat C_{33}$.  The eigenvalue of $\hat H_1$ acting on the
basis state $\vert(\lambda,\mu)\nu;I\rangle$ is $2\nu_1-\nu_2-\nu_3=3\nu_1-(\lambda+2 \mu)$ and is thus fully specified by $\nu_1$.

The highest weight state is of the form (see \cite{rowe1999representations}):
\beq
\ket{(\lambda,\mu)\hbox{hw}}= \vert (\lambda,\mu)\lambda+\mu,\mu,0;\textstyle\frac{1}{2}\mu \rangle\, .
\label{eq:highestweightstate}
\eeq 
All states of an $\sugroup(2)\oplus \mathfrak{u}(1)$ multiplet have the same occupation number $\nu_1$, but there may be more than one multiplet with the same
value of $\nu_1$ when some weights are degenerate. 

\begin{figure}[h!]
\centering
\includegraphics[scale=0.35]{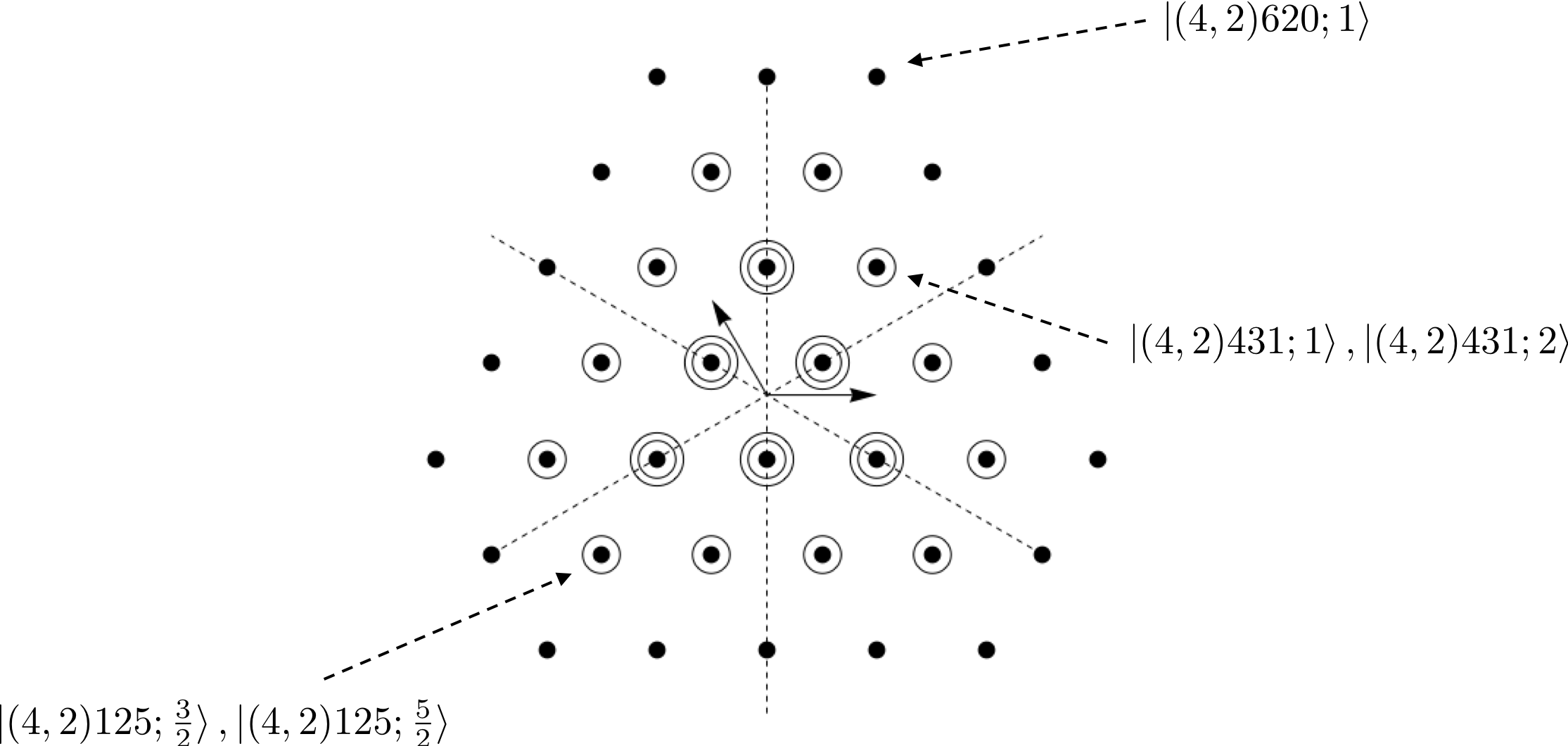}
\caption{The weight diagram for the irrep $(4,2)$, showing also the location of some states $\vert (4,2)\nu_1\nu_2\nu_3;I\rangle $ on the weight diagram.}
\label{fig:42weightdiagram}
\end{figure}

An example of the weight diagram for the irrep $(4,2)$, along with the position of selected states on the diagram, their occupation numbers and 
$\sugroup(2)$ label $\hdg{I}$,
is given in Figure \ref{fig:42weightdiagram}.  Dots represent weights with multiplicity $1$, circled dots are weights with multiplicity $2$, double-circled
dots are weights with multiplicity $3$.

\subsection{Some $\sugroup(2)$ tensors and a projection}

Let $L=\frac{1}{2}p$ and define the components of the  $\sugroup(2)\oplus u(1)$ tensor $\hat T^L_M$ 
\cite{rowe1997algebraic} by 
\beqa
\hat T^L_M&=&\cg{\frac{1}{2}(L+M)}{\frac{1}{2}(L+M)}{\frac{1}{2}(L-M)}{-\frac{1}{2}(L-M)}{L}{M}
\textstyle\frac{(2L)!}{(L+M)!(L-M)!}\,(\hat C_{31})^{L-M} (\hat C_{21})^{L+M} \\
&=& \sqrt{\frac{(2L)!}{(L+M)!(L-M)!}}\,(\hat C_{31})^{L-M} (\hat C_{21})^{L+M}  \label{eq:tensorLM}
\eeqa
so that, for instance, we have
$$
{\renewcommand{\arraystretch}{1.65}
\begin{array}{ccl ||ccl}
L&M&\hat T^L_M&L&M&\hat T^L_M\\
\hline
1&1&\hat C_{21}^2&3/2&3/2&\hat C_{21}^3\\
1&0&\sqrt{2}\hat C_{31}\hat C_{21}&3/2&1/2& \sqrt{3}\hat C_{31}\hat C_{21}^2\\
1&-1&\hat C_{31}^2&3/2&-1/2&\sqrt{3}\hat C_{31}^2\hat C_{21}\\
&&&3/2&-3/2&\hat C_{31}^3
\end{array}}
$$
Note that these operators also decrease the $\mathfrak{u}(1)$ label $\nu_1$ by $2L$.

With this, for fixed $J$ and $M_J$, we can use the Wigner-Eckart theorem to note that %the sum
\beqa
&&\sum_{M_I} \cg{I}{M_I}{L}{M_J-M_I}{J}{M_J}\,\hat T^{L}_M\ket{(\lambda,\mu)\eta_1\eta_2\eta_3;I }\nonumber \\
&&\  = \ket{(\lambda,\mu)\eta_1-2L,\eta_2+L+M_J-M_I,\eta_3+L-M_J+M_I;J } \nonumber \\
&&\qquad\times \frac{\rme{(\lambda,\mu)\eta_1-2 L;J}{\hat T^L}{(\lambda,\mu)\eta_1;I} }{\sqrt{2J+1}}
\label{shiftedexpression}
\eeqa
where $M_I=\hf(\eta_2-\eta_3)$,  $J$ is one of $I+L, I+L-1, \ldots, \vert I-L\vert$, and 
$\rme{(\lambda,\mu)\eta_1-2 L;J}{\hat T^L}{(\lambda,\mu)\eta_1;I}$ is a reduced matrix element, which depends on the $\mathfrak{u}(1)$ 
labels $\eta_1-2L$ and $\eta_1$, and on the $\mathfrak{su}(2)$ labels $J$ and $I$.

\begin{figure}[h!]
\centering
\includegraphics[scale=0.5]{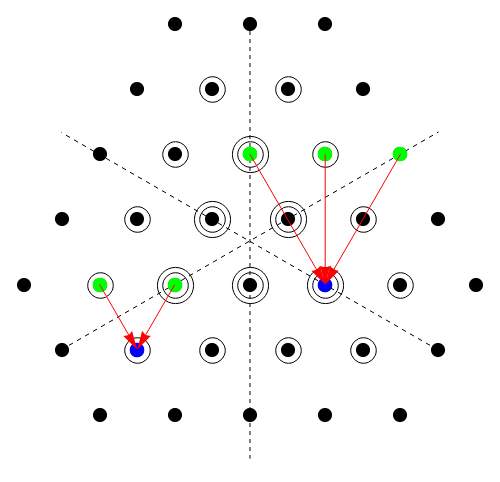}
\caption{Two examples of lowering tensors.  A linear combination of the two components of the 
$\hat{T}^{\frac{1}{2}}$ tensor acting on two different states produces
a single state in the layer directly below; a linear combination of the three components of the $T^1$ tensor, each acting on different states, produces
a state two layers below.}
\label{fig:tensoraction}
\end{figure}

Thus, the action of the \emph{linear combination} of operators $\hat T^{L}_M$  ($M=M_J-M_I$) operators acting on 
states in the same
$\mathfrak{su}(2)\oplus \mathfrak{u}(1)$ multiplet
will shift these states down by $p$ layers ({\it i.e.} decrease the $\mathfrak{u}(1)$ label by $2p$) on the weight diagram to a \emph{specific state} 
in a \emph{specific multiplet}; this is in contradistinction to the action of \emph{specific} operator 
on a single state that will
generically produce a \emph{linear combination} of states.  This idea is illustrated in Fig.\ref{fig:tensoraction}.

\subsection{Closed form expression for  $\rme{(\lambda,\mu)\lambda+\mu-p;J}{\hat T^{\frac{1}{2}p}}{(\lambda,\mu)\lambda+\mu;\hf\mu}$}

We start by examining the action of $\hat{T}^L_M$ on states $\ket{(\lambda,\mu)\lambda+\mu,\mu-s,s;\hf\mu}$ 
 in the same multiplet as the highest weight state.  
 All these states have $\nu_1= \lambda+\mu$.  

One can easily obtain the reduced matrix element of the tensor operator $\hat T^{\frac{1}{2}p}$ acting these states.
% of the form
%$\ket{(\Lambda,\mu)\Lambda+\mu,\mu-s,s;\hf\mu}$ which are in the same 
%$\mathfrak{su}(2)_{23}\oplus \mathfrak{u}(1)$ multiplet as the highest weight state.
First choose $M_J=\hf(\mu-p)$ so that
\beq
\hf\mu-s+M=\hf(\mu-p) \qquad\Rightarrow \qquad M=s-\hf p\, .
\label{Mcondition}
\eeq
from which
\beqa
\kern-2em&&\sum_{s=0}^p \cg{\frac{1}{2}\mu}{\frac{1}{2}\mu-s}{\frac{1}{2}p}{s-\frac{1}{2} p}{J}{\hf(\mu-p)}\,
\hat T^{\frac{1}{2} p}_{s-\frac{1}{2} p}\ket{(\lambda,\mu)\lambda+\mu,\mu-s,s;\hf\mu }\nonumber \\
\kern-2em&& = 
\ket{(\lambda,\mu)\lambda+\mu-p,\mu,p;J}
\frac{\rme{(\lambda,\mu)\lambda+\mu-p;J}{\hat T^{\frac{1}{2} p}}{(\lambda,\mu)\lambda+\mu;\hf \mu}}{\sqrt{2J+1}}
\label{generalshift}
\eeqa

Next, compute 
\beqa
\kern-1em
&&\bra{(\lambda,\mu)\lambda+\mu-p,\mu-s,s+p;J}P_{12}(\hat C_{32})^pP_{12}\ket{(\lambda,\mu)\lambda+\mu,\mu-s,s;\textstyle\frac{1}{2}\mu} 
\nonumber \\
\kern-1em
&&=\bra{(\lambda,\mu)\lambda+\mu-p,\mu-s,s+p;J}(\hat C_{31})^p\ket{(\lambda,\mu)\lambda+\mu,\mu-s,s;\textstyle\frac{1}{2}\mu} 
\eeqa
where $P_{12}$ is the permutation with matrix element given by  \cite{rowe1999representations} in terms of a $6j$-symbol:
\beqa
&&\bra{(\lambda,\mu)\nu_2 \nu_1\nu_3 ;I'} P_{12}\ket{(\lambda,\mu)\nu_1\nu_2\nu_3;I} 
\nonumber \\
&&\quad =(-1)^{\frac{1}{2}(\nu_3-\lambda)-(I+I')+\mu}
\sqrt{(2I+1)(2I'+1)}
{\renewcommand{\arraycolsep}{1.75pt}
\renewcommand{\arraystretch}{1.5}
\sixj{\frac{1}{2}\nu_1}{ \frac{1}{2}\nu_3}{I'}{ \frac{1}{2}\nu_2}{\frac{1}{2}\lambda }{I}
}\, .
\eeqa
From this we obtain
\beqa
&&\rme{(\lambda,\mu)\lambda+\mu-p;J}{\hat T^{\frac{1}{2} p}}{(\lambda,\mu)\lambda+\mu;\hf\mu} \nonumber \\
&&\quad =
 \displaystyle\sqrt{\frac{(2J+1)(\lambda+\mu+1)!(J+\hf (\mu+p)+1)!(\hf(\mu+p)-J)!}{(\lambda+\mu-p)!\mu!}}\nonumber \\
 &&\qquad\times (-1)^{(p-2J-2\lambda-\mu)/2}
{\renewcommand{\arraystretch}{1.25} \sixj{\hf(\lambda+\mu-p)}{\hf p}{\hf(\lambda+\mu)}{\hf \mu}{\hf \lambda}{J}}
\eeqa

The $6j$ symbol has a closed form expression \cite{varshalovich1988quantum} 
since the first and second entry of the first row sum to the third.  As a result 
%of this simplification 
we reach the final expression
\beqa
\kern-1em&&\rme{(\lambda,\mu)\lambda+\mu-p;J}{\hat T^{\frac{1}{2}p}}{(\lambda,\mu)\lambda+\mu;\hf\mu} \nonumber \\
\kern-1em&& =(-1)^{(p-2J+\mu)/2}
\sqrt{\displaystyle\frac{(2J+1)(\lambda+\mu+1)!\lambda!p!}{(\lambda-J+\frac{1}{2}(\mu-p))!(\lambda+J+\frac{1}{2}(\mu-p)+1)!}}\, .
\label{rmegeneral}
\eeqa
%%%

\subsection{General matrix element}

For the action of $\hat T^k$ on states $\ket{(\lambda,\mu)\nu;I}$  
not in the same multiplet as the highest weight state, we can still obtain a reduced matrix element 
starting with
\beqa 
&&\rme{(\lambda,\mu)\lambda+\mu-p-2k;J'}{\hat T^{k}}{(\lambda,\mu)\lambda+\mu-p;J}\nonumber \\
&&\quad=\sqrt{2J'+1}\sum_{m_kM_J}\cg{J}{M_J}{k}{m_k}{J'}{M'}\nonumber \\
&&\quad \times \bra{(\lambda,\mu)\lambda+\mu-p-2k,\hf(\mu+p)+k+M',\hf(\mu+p)+k-M';J'}\nonumber \\
&&\quad\qquad \hat T^k_{m_k}
\ket{(\lambda,\mu)\lambda+\mu-p,\hf(\mu+p)+M_J,\hf(\mu+p)-M_J;J}
\eeqa 
and writing the state $\ket{(\lambda,\mu)\lambda+\mu-p,\hf(\mu+p)+M_J,\hf(\mu+p)-M_J;J}$ as
\beqa 
&&\ket{(\lambda,\mu)\lambda+\mu-p,\hf(\mu+p)+M_J,\hf(\mu+p)-M_J;J}\nonumber \\
&&\; =
\displaystyle\frac{\sqrt{2J+1}}{\rme{(\lambda,\mu)\lambda+\mu-p}{\hat T^{\frac{1}{2}p}}{(\lambda,\mu)\lambda+\mu;\frac{1}{2}\mu}}\nonumber \\
&& \;\times 
\sum_{m_sm_p}\cg{\frac{1}{2}\mu}{m_s}{\frac{1}{2} p}{m_p}{J}{M_J}\hat T^{\frac{1}{2}p}_{m_p}
\ket{(\lambda,\mu)\lambda+\mu,\hf\mu+m_s,\hf\mu-m_s;\frac{1}{2}\mu}\, .
\eeqa 
One can then combine the tensors
\beq 
\hat T^k_{m_k}\hat T^L_M=\sum_{\bar M}\cg{k}{m_k}{L}{M}{k+L}{\bar M}T^{k+L}_{\bar M}
\eeq 
to obtain, after straightforward manipulations, the expression
\beqa 
&&\rme{(\lambda,\mu)\lambda+\mu-p-2k;J'}{\hat T^k}{(\lambda,\mu)\lambda+\mu-p;J}\nonumber \\
&&=(2J+1)\sqrt{2k+p+1}(-1)^{\frac{1}{2}(\mu+p)+k+J'}\sixj{k}{\frac{1}{2}p}{k+\frac{1}{2}p}{\frac{1}{2}\mu}{J'}{J}\nonumber \\
&&\, \times\frac{
\rme{(\lambda,\mu)\lambda+\mu-p-2k;J'}{\hat T^{k+\frac{1}{2}p}}{(\lambda,\mu)\lambda+\mu;\frac{1}{2}\mu}}{\rme{(\lambda,\mu)\lambda+\mu-p;J}
{\hat T^{\frac{1}{2}p}}{(\lambda,\mu)\lambda+\mu;\frac{1}{2}\mu}}\, .\label{eq:kpowertensor}
\eeqa 
Once again, the $6j$ symbol has a closed form expression and 
the ratio of reduced matrix elements on the right hand side can be evaluated using 
Eq.(\ref{rmegeneral}).

The special case $k=\frac{1}{2}$ produces the reduced matrix elements for the generators $\hat C_{31}$ and $\hat C_{21}$ as components of the $\hat T^{\frac{1}{2}}$ tensor:
\beqa
&&\rme{(\lambda,\mu)\lambda+\mu-(p+1);J}{\hat T^{\frac{1}{2}}}{(\lambda,\mu)\lambda+\mu-p;I}\nonumber \\
&&\quad=(-1)^{(\mu+p+1)/2+J}\,(2I+1)\sqrt{p+2}\displaystyle\sixj{\frac{1}{2}}{\hf p}{\hf(p+1)}{\hf \mu}{J}{I}\nonumber \\
&&\quad\quad  \times 
\frac{\rme{(\lambda,\mu)\lambda+\mu-(p+1);J}{\hat T^{(p+1)/2}}{(\lambda,\mu)\lambda+\mu;\hf\mu}}
{\rme{(\lambda,\mu)\lambda+\mu-p;I}{\hat T^{p/2}}{(\lambda,\mu)\lambda+\mu;\hf\mu}} \, .
\eeqa

To obtain the matrix elements of the raising operators 
$\hat C_{13}$ and $\hat C_{12}$, we note those are components of the tensor $\bar T^{\frac{1}{2}}$ 
adjoint to
$\hat T^{\frac{1}{2}}$, with 
\beq
\bar T^{\mysmallfrac{1}{2}}_{\mysmallfrac{1}{2}}= (\hat T^{\mysmallfrac{1}{2}}_{\mysmallfrac{-1}{2}})^\dagger = \hat C_{13}\, ,\qquad 
\bar T^{\mysmallfrac{1}{2}}_{\mysmallfrac{-1}{2}}=- (\hat T^{\mysmallfrac{1}{2}}_{\mysmallfrac{1}{2}})^\dagger = -\hat C_{12}
\eeq
and 
\beqa
&&\rme{(\lambda,\mu)\lambda+\mu-(p-1);J}{\bar T^{\frac{1}{2}}}{(\lambda,\mu)\lambda+\mu-p;I} \nonumber \\ 
&&\, =
(-1)^{I+\frac{1}{2}-J}\, \rme{(\lambda,\mu)\lambda+\mu-p;I}{\hat T^{\frac{1}{2}}}{(\lambda,\mu)\lambda+\mu-(p-1);J}\, .
\eeqa

Some specialized expressions are provided in \ref{sec:specialcase}

\section{Constructing SU(3) Clebsch-Gordan coefficients}

\hdg{In this section we deal with SU(3) CG.  We first obtain a recursion relation satisfied by the CG coefficients
needed for the highest weight state of $(p_2,q_2)$ in $(p_1,q_1)\otimes (\lambda,\mu)$.  This recursion relation
does not depend on the number of copies of $(p_2,q_2)$ in the decomposition so is valid quite generally, although
it can be solved in terms of a single ``seed'' coefficient when $(p_2,q_2)$ occurs only once. 
We then show how}, once the highest weight state has been obtained through a recursion relation, we can obtain the full 
CG through a 
summation that involves matrix elements of  tensors obtained in Eq.(\ref{eq:tensorLM}), and a $9j$ symbol.  

We denote by $\cg{(p_1,q_1)}{\nu;J}{(\lambda,\mu)}{n, I_n}{(p_2,q_2)}{\bar\nu;\bar I}$ 
 the Clebch-Gordan coefficient that occurs in the construction of the state $\vert (p_2,q_2)\bar \nu;\bar I\rangle $ from the states
 $\vert (p_1,q_1) \nu;J\rangle$ and  $\vert (\lambda,\mu) n; I_n\rangle$.  If the irrep $(p_2,q_2)$ occurs more than once in the
 decomposition of $(p_1,q_1)\otimes (\lambda,\mu)$, we assume we have identified a specific copy of $(p_2,q_2)$ for which we wish
 to obtain the CGs.

\subsection{Two recursion relations for the highest weight state}

For the highest weight state of $(p_2,q_2)$ write
\beqa
&&\ket{(p_2,q_2)\hbox{hw}}\nonumber \\
&&\, =
\sum_{\nu J (n)I_n}\cg{(p_1,q_1)}{\nu;J}{(\lambda,\mu)}{n I_n}{(p_2,q_2)}{p_2+q_2,q_2,0;\frac{1}{2}q_2}
\ket{(p_1,q_1)\nu; J}\ket{(\lambda,\mu)n; I_n}\, . \label{generalhws}
\eeqa

The indices $\nu$ and $n$ must satisfy the following constraints: the total number of excitations in the composite system is
$p_1+2q_1+\lambda+2\mu$, which ought to equal $p_2+2q_2+3k$.  The integer $k$ satisfies
\beq
k=\frac{1}{3}\left( p_1 +\lambda -p_2+2( q_1+\mu -q_2)\right) \label{eq:kdefinition}
\eeq 
and is such that
\beqa
\nu_1+n_1=p_2+q_2+k\, ,\quad 
\nu_2+n_2=q_2+k\, ,\quad
\nu_3+n_3=k\, . 
\eeqa
The index $k$ is required because the highest weight of $(p_2,q_2)$ can be multiplied by the $k$'th power of the $3\times 3$ determinant
\beq
\left\vert\begin{array}{ccc} 
\hat a_{11}^\dagger & \hat a_{12}^\dagger &\hat a_{13}^\dagger\\
\hat a_{21}^\dagger & \hat a_{22}^\dagger &\hat a_{23}^\dagger\\
\hat a_{31}^\dagger &\hat a_{32}^\dagger &\hat a_{33}^\dagger
\end{array}\right\vert^k\, , \label{eq:determinantfactor}
\eeq
and this determinant transforms by the $\sugroup{(3)}$ irrep $(0,0)$ but 
add $3k$ excitations to the system without changing the irrep label
$(p_2,q_2)$.  The requirement
\beq
p_1+2q_1+\lambda+2\mu= p_2+2q_2+3k \label{def:grade}
\eeq
guarantees that the number of excitations in the separate systems is the same as that in the resulting irrep.
The label $k$ can be ignored when 
constructing the CGs but it does play a role 
in analyzing  symmetries of the CGs, as discussed in Secs \ref{sec:Weylsymmetries} and \ref{sec:additional};
it also connects the $\mathfrak{u}(1)$ labels in $(p_1,q_1)\otimes (\lambda,\mu)$ and $(p_2,q_2)$. 
This index $k$ is called the {\it grade} 
in \cite{rowe1997algebraic}, and identifying $k$ as a difference in the number of quanta through Eq.(\ref{def:grade}) 
provides a physical interpretation to this quantity.

By definition of the highest weight state
\beqa
\kern-2em&&0=\hat C_{12}\ket{(p_2,q_2)\hbox{hw}}\, ,\\
\kern-2em&&= 
\sum_{\nu J (n)I_n}\cg{(p_1,q_1)}{\nu;J}{(\lambda,\mu)}{n I_n}{(p_2,q_2)}{p_2+q_2,q_2,0;\frac{1}{2}q_2}
\left[C_{12}\ket{(p_1,q_1)\nu;J}\right] \ket{(\lambda,\mu)nI_n} \nonumber \\
\kern-2em&&+\sum_{\nu J (n)I_n}\cg{(p_1,q_1)}{\nu;J}{(\lambda,\mu)}{n I_n}{(p_2,q_2)}{p_2+q_2,q_2,0;\frac{1}{2}q_2}
\ket{(p_1,q_1)\nu;J}\left[C_{12}\ket{(\lambda,\mu)nI_n}\right]  \, .
\eeqa
Closing with $\bra{(p_1,q_1)\nu'J'}\bra{(\lambda,\mu)n'{I'}_n}$ we get the basic recursion relation
\beqa
&&0=\sum_{(\nu) J}\cg{(p_1,q_1)}{\nu;J}{(\lambda,\mu)}{n' I'_n}{(p_2,q_2)}{p_2+q_2,q_2,0;\frac{1}{2}q_2}
\bra{(p_1,q_1)\nu;J}\hat C_{21}\ket{(p_1,q_1)\nu'I'}  \nonumber \\
&&+\sum_{(n)I_n}\cg{(p_1,q_1)}{\nu';J'}{(\lambda,\mu)}{n I_n}{(p_2,q_2)}{p_2+q_2,q_2,0;\frac{1}{2}q_2}
\bra{(\lambda,\mu)nI_n}\hat C_{21}\ket{(\lambda,\mu)n'{I'}_n} \, . \label{basicCGrecursion}
\eeqa
We can rewrite the recursion for the highest weight state as
\beqa
0&=&\sum_{(\nu_1)m_\nu J}
\rcg{(p_1,q_1)}{\nu_1;J}{(\lambda,\mu)}{n'_1 I'_n}{(p_2,q_2)}{p_2+q_2;\frac{1}{2}q_2}
\cg{J}{m_{\nu}}{I'_n}{m'_n}{\frac{1}{2}q_2}{\frac{1}{2}q_2}\nonumber \\
&&\qquad \times \frac{\rme{(p_1,q_1)\nu_1;J}{\hat T^{\frac{1}{2}}}{(p_1,q_1)\nu'_1;J'}}{\sqrt{2J+1}}
\cg{J'}{m'_{\nu}}{\frac{1}{2}}{\frac{1}{2}}{J}{m_\nu} 
\nonumber \\
&+&
\sum_{(n_1)I_nm_n}\rcg{(p_1,q_1)}{\nu'_1;J'}{(\lambda,\mu)}{n_1 I_n}{(p_2,q_2)}{p_2+q_2;\frac{1}{2}q_2}
\cg{J'}{m'_\nu}{I_n}{m_n}{\frac{1}{2}q_2}{\frac{1}{2}q_2}\nonumber \\
&&\qquad\times \frac{\rme{(\lambda,\mu)n_1I_n}{\hat T^{\frac{1}{2}}}{(\lambda,\mu)n_1'{I'}_n}}{\sqrt{2I_n+1}}
\cg{I_n}{m_n}{\frac{1}{2}}{\frac{1}{2}}{I'}{m_n'}  \, . \label{recursiononehalf}
\eeqa
where $m_{\nu}=\frac{1}{2}(\nu_2-\nu_3), m_n=\frac{1}{2}(n_2-n_3)$ \emph{etc}.  \hdg{Here, we have introduced the usual factorization
of an $\mathfrak{su}(3)$ CG coefficient into a {\it reduced} (or doubled-barred) coefficient and an $\mathfrak{su}(2)$ CG:
\beq
\cg{(p_1,q_1)}{\nu ;J}{(\lambda,\mu)}{n';I'_n}{(p_2,q_2)}{\mu;I}:=
\rcg{(p_1,q_1)}{\nu_1;J}{(\lambda,\mu)}{n'_1;I'_n}{(p_2,q_2)}{\mu_1;I}
\cg{J}{m_\nu}{I'_n}{m'_n}{I}{m_\mu}\, .
\eeq 
}

A similar recursion can be obtained starting with $\hat C_{13}$, so that we have a generalized form Eq.(\ref{recursiononehalf}) as
\beqa
0&=&\sum_{(\nu_1)J}
\frac{\rme{(p_1,q_1)\nu_1;J}{\hat T^{\frac{1}{2}}}{(p_1,q_1)\nu'_1;J'}}{\sqrt{2J+1}}
\rcg{(p_1,q_1)}{\nu_1;J}{(\lambda,\mu)}{n'_1 I'_n}{(p_2,q_2)}{p_2+q_2;\frac{1}{2}q_2}
\nonumber \\
&&\qquad \times \sum_{m_\nu}
\cg{J}{m_\nu}{I'_n}{m'_n}{\frac{1}{2}\hdg{q_2}}{\frac{1}{2}q_2}
\cg{J'}{m'_\nu}{\frac{1}{2}}{k}{J}{m_\nu} 
\nonumber \\
&+&
\sum_{(n_1)I_n}
\frac{\rme{(\lambda,\mu)n_1I_n}{\hat T^{\frac{1}{2}}}{(\lambda,\mu)n_1'{I'}_n}}{\sqrt{2I_n+1}}
\rcg{(p_1,q_1)}{\nu'_1;J'}{(\lambda,\mu)}{n_1 I_n}{(p_2,q_2)}{p_2+q_2;\frac{1}{2}q_2}
\nonumber \\
&&\qquad\times \sum_{m_n}\cg{J'}{m'_\nu}{I_n}{m_n}{\frac{1}{2}\hdg{q_2}}{\frac{1}{2}q_2}
\cg{I_n}{m_n}{\frac{1}{2}}{\frac{1}{2}}{I'_n}{m'_n}  \, . \label{recursionk}
\eeqa

Multiplication by $\cg{J'}{m'_\nu}{\frac{1}{2}}{m}{\tilde J}{\tilde m_\nu}$ followed by summation over $\hdg{m}'_\nu$ and $m$ produces, upon using
the orthogonality property of CGs under summation,
a simplification of Eq.(\ref{recursionk}) in the form
\beqa
0&=&\sum_{(\nu_1)}\frac{\rme{(p_1,q_1)\nu_1\tilde J}{\hat T^{\frac{1}{2}}}{(p_1,q_1)\nu'_1J'}}{\sqrt{2\tilde J+1}}
\rcg{(p_1,q_1)}{\nu_1\tilde J}{(\lambda,\mu)}{n'_1I'_n}{(p_2,q_2)}{p_2+q_2;\frac{1}{2}q_2}\nonumber \\
&&\times \cg{\tilde J}{\tilde m_\nu}{I'_n}{k}{\frac{1}{2}\hdg{q_2}}{\frac{1}{2}q_2}\nonumber\\
&+&\sum_{(n_1)I_n}\frac{\rme{(\lambda,\mu)n_1I_n}{\hat T^{\frac{1}{2}}}{(\lambda,\mu)n'_1I'_n}}{\sqrt{2I_n+1}}
\rcg{(p_1,q_1)}{\nu'_1\tilde J'}{(\lambda,\mu)}{n_1I_n}{(p_2,q_2)}{p_2+q_2;\frac{1}{2}q_2}\nonumber\\
&&\times\sum_{k m'_\nu m_n}
\cg{I'_n}{m'_n}{\frac{1}{2}}{k}{I_n}{m_n}
\cg{J'}{m'_\nu}{I_n}{m_n}{\frac{1}{2}q_2}{\frac{1}{2}q_2}
\cg{J'}{m'_\nu}{\frac{1}{2}}{k}{\tilde J}{\tilde m_\nu}
\, .
\eeqa
The arguments in the product of the three CGs can be rearranged to an expression containing an $\sugroup{(2)}$ $6j$-symbol \cite{varshalovich1988quantum}
leading to a direct recursion for the reduced CG:
\beqa
&&\rcg{(p_1,q_1)}{\nu'_1-1;\tilde J}{(\lambda,\mu)}{n'_1;I'_n}{(p_2,q_2)}{p_2+q_2;\frac{1}{2}q_2}\nonumber \\
&&=(2\tilde J+1)\sum_{I_n}
\frac{\rme{(\lambda,\mu)n'_1-1;I_n}{\hat T^{\frac{1}{2}}}{(\lambda,\mu)n'_1I'_n}}{\rme{(p_1,q_1)\nu'_1-1;\tilde J}{\hat T^{\frac{1}{2}}}{(p_1,q_1)\nu'_1;J'}} (-1)^{I_n+J'+\frac{1}{2}q_2+1}
\nonumber \\
%%%%
&&\times % (-1)^{I_n+J'+q_2/2+1}
\rcg{(p_1,q_1)}{\nu'_1J'}{(\lambda,\mu)}{n'_1-1;I_n}{(p_2,q_2)}{p_2+q_2;\frac{1}{2}q_2}
\sixj{\frac{1}{2}}{J'}{\tilde J}{\frac{1}{2}q_2}{I'_n}{I_n}\, . \label{recursion_reduced} %tilde J changed to q/2
\eeqa
Similar manipulations can also yield a slightly different recursion:
\beqa
&&\rcg{(p_1,q_1)}{\nu'_1;J'}{(\lambda,\mu)}{n'_1-1;\tilde{I}_n}{(p_2,q_2)}{p_2+q_2;\frac{1}{2}q_2}\nonumber \\
&&=(2\tilde I_n+1)\sum_{J}
\frac{\rme{(p_1,q_1)\nu'_1-1;J}{\hat T^{\frac{1}{2}}}{(p_1,q_1)\nu'_1;J'}}
{\rme{(\lambda,\mu)n'_1-1;\tilde{I}_n}{\hat T^{\frac{1}{2}}}{(\lambda,\mu)n'_1I'_n}} %(-1)^{J'-\tilde I_n+\frac{1}{2}q_2+2I'_n}
 \nonumber \\
&&\kern-2em \times  (-1)^{J'-\tilde I_n+\frac{1}{2}q_2+2I'_n}
\rcg{(p_1,q_1)}{\nu'_1-1;J\ }{(\lambda,\mu)}{n'_1;I'_n}{(p_2,q_2)}{p_2+q_2;\frac{1}{2}q_2}
\sixj{\frac{1}{2}}{J'}{J}{\frac{1}{2}q_2}{I'_n}{\tilde I_n}\, . \label{recursion_reduced2}
\eeqa
Note that in Eqs.(\ref{recursion_reduced}) and (\ref{recursion_reduced2}), the steps in the angular momenta are at most
$J'=J\pm \frac{1}{2}$ and $I'_n=\tilde I_n\pm \frac{1}{2}$, \emph{i.e.} the right hand side of the recursion 
of Eqs.(\ref{recursion_reduced}) and (\ref{recursion_reduced2}) contain at most two terms.   
This can be compared with \cite{williams1996su3}.  \hdg{If the multiplicity of $(p_2,q_2)$ in the tensor product
is $1$, every reduced CG can be recursively determined in terms of a single ``seed'' coefficient.  If the multiplicity
 is $\sigma$, the recursion leads to a final expression involving a general linear combination of $\sigma$ coefficients subject
 only to a single normalization constraint; in other words there is more than one possible highest weight (unsurprisingly!), 
 and there is no $\mathfrak{su}(3)$ argument to further specify each highest weight state: one must resort to other symmetries or arguments to fix the linear combinations.} 

The recursion of Eq.(\ref{basicCGrecursion}) clearly does not depend on the ordering of
$(p_1,q_1)$ and $(\lambda,\mu)$.  It follows that the CGs
$\cg{(p_1,q_1)}{\nu;J}{(\lambda,\mu)}{n' I'_n}{(p_2,q_2)}{p_2+q_2,q_2,0;\frac{1}{2}q_2}$ and 
$\cg{(\lambda,\mu)}{n' I'_n}{(p_1,q_1)}{\nu;J}{(p_2,q_2)}{p_2+q_2,q_2,0;\frac{1}{2}q_2}$ satisfy the same recursion relations.
They thus differ by at most a phase, which depends on the seed coefficient of the recursion.  The phase convention we use 
\cite{de1963octet} is to
take 
\beq
\cg{(p_1,q_1)}{\hbox{hw}}{(\lambda,\mu)}{n' \tilde{I}'_n}{(p_2,q_2)}{\hbox{hw}}\ge 0\, .  \label{eq:phaseconvention}
\eeq
where $\tilde{I}'_n$ is the largest value of $I'_n$ compatible with $n'$.

This is a convenient place to expand on this phase convention.  The construction of 
Eq.(\ref{eq:fullsu3states}) 
contains within it an implicit choice of phases since the ordering 
of the \hdg{terms in the products} is chosen by the definition of the states.    
It turns out that this implicit convention is not completely compatible with the choice of Eq.(\ref{eq:phaseconvention}).  This is 
best illustrated with an example: the highest weight state of the irrep $(0,1)$ constructed according to Eq.(\ref{eq:fullsu3states})  is
\beq
\ket{(0,1)210;\textstyle\frac{1}{2}}=-\frac{1}{\sqrt{2}}\left(\hat a^\dagger_{11}\hat a^\dagger_{22}-\hat a^\dagger_{12}\hat a^\dagger_{21}\right)\ket{0} \label{eq:directstate}
\eeq 
and can easily be identified with the linear combination
\beq 
-\frac{1}{\sqrt{2}}\ket{(1,0)100}_1\ket{(1,0)010}_2 + \frac{1}{\sqrt{2}}\ket{(1,0)010}_1\ket{(1,0)100}_2\, ,
\eeq 
which does not satisfy the convention of Eq.(\ref{eq:phaseconvention}).
%This will be discussed at length in a specific example.

Although this phase mismatch is unfortunate,  there are good reasons to retain Eq.(\ref{eq:phaseconvention}) since ${SU}(3)$ group functions \cite{rowe1999representations} can be
used to obtain CGs.  
If the irrep $(p_2,q_2)$ 
occurs once in the decomposition of $(p_1,q_1)\otimes (\lambda,\mu)$, one can evaluate the integral
\beqa
&&\int d\Omega D^{(p_1,q_1)}_{\hbox{\tiny{hw}};\hbox{\tiny{hw}}}(\Omega)\
D^{(\lambda,\mu)}_{\small{n'{I'_n}};\small{n'{I'_n}}}(\Omega)\
\left(D^{(p_2,q_2)}_{\hbox{\tiny{hw}};\hbox{\tiny{hw}}}(\Omega)\right)^* \nonumber \\
&&\qquad =\frac{\hbox{vol}}{\hbox{dim}(p_2,q_2)}
\Bigl\vert\cg{(p_1,q_1)}{\hbox{hw}}{(\lambda,\mu)}{n'{I}'_n}{(p_2,q_2)}{\hbox{hw}}\Bigr\vert^2
\label{eq:integralform}
\eeqa
\hdg{where $n'=(n'_1n'_2 n'_3)$ satisfies
\beqa 
p_1+q_1+ n'_1&= &p_2+q_2+k\, ,\\
q_1+n'_2&= &q_2+k\, ,\\
n'_3&=&k\, ,
\eeqa 
$I'_n=\frac{1}{2}(\lambda-n'_1)$, and \cite{de2018simple}\cite{russell2017direct}\cite{murnaghan1962unitary}}
\beq
\kern-5em d\Omega=\sin\beta_1\,\cos\frac{1}{2}\beta_2\left(\sin\frac{1}{2}\beta_2\right)^3\,\sin\beta_3\,d\alpha_1\,d\alpha_2\,d\alpha_3
\,d\gamma_1\,d\gamma_2\,d\beta_1\,d\beta_2\,d\beta_3. 
\eeq   
The integral can be evaluated analytically
since functions of the type $D^{(p_1,q_1)}_{\hbox{\tiny{hw}};\hbox{\tiny{hw}}}(\Omega)$ have a fairly simple form:
\beqa
D^{(p,q)}_{\hbox{\tiny{hw}};\hbox{\tiny{hw}}}(\Omega)
   &=&\left(D^{(1,0)}_{(100)0;(100)0}(\Omega)\right)^p \,\left(D^{(0,1)}_{(110)\frac{1}{2};(110)\frac{1}{2}}(\Omega)\right)^q\, . 
\eeqa
where
\beqa
&&\,\, D^{(1,0)}_{(100)0;(100)0}(\Omega)=e^{-2 i\gamma_1} (\cos \left(\textstyle\frac{1}{2}\beta_2\right))\, \\
&&D^{(0,1)}_{(110)\frac{1}{2};(110)\frac{1}{2}}(\Omega)
= e^{-\frac{1}{2} i  (2 \alpha_1+2 \gamma_1+\gamma_2)} \nonumber \\
&& \times \left(e^{i \alpha_1} \cos
   \left(\textstyle\frac{1}{2}\beta_1\right) \cos \left(\textstyle\frac{1}{2}\beta_3\right)-e^{i \alpha_3} \sin
   \left(\textstyle\frac{1}{2}\beta_1\right) \cos \left(\textstyle\frac{1}{2}\beta_2\right) \sin \left(\textstyle\frac{1}{2}\beta_3\right)\right) .
\eeqa

\subsection{General expression for the CG using $9j$ symbols}

Suppose we are to construct the CG coefficients for the \hdg{direct product} $(p_1,q_1)\otimes (\lambda,\mu)\to (p_2,q_2)$
for $(p_2,q_2)$ is fixed, i.e. if the irrep $(p_2,q_2)$ occurs more than once in the decomposition $(p_1,q_1)\otimes (\lambda,\mu)$ 
then we have selected a particular copy.  

We suppose we are given the highest weight state
\beqa 
&&\ket{(p_2,q_2)\hbox{hw}}\nonumber \\
&&=
\sum_{\nu J (n)I_n}\cg{(p_1,q_1)}{\nu;J}{(\lambda,\mu)}{n I_n}{(p_2,q_2)}{p_2+q_2,q_2,0;\frac{1}{2}q_2}
\ket{(p_1,q_1)\nu;J}\ket{(\lambda,\mu)n;I_n}\, , 
\eeqa
of Eq.(\ref{generalhws}) with explicit knowledge of CGs $\cg{(p_1,q_1)}{\nu;J}{(\lambda,\mu)}{n I_n}{(p_2,q_2)}{p_2+q_2,q_2,0;\frac{1}{2}q_2}$.
These can be obtained using the recursion of Eq.(\ref{basicCGrecursion}) or any other means.

We identify a target state $\ket{(p_2,q_2)p_2+q_2-p,\hf (q_2+p)+M_I,\hf (q_2+p)-M_I;I}$.  
From this
\beqa
&&\ket{(p_2,q_2)p_2+q_2-p,\textstyle\frac{1}{2}(p_2+q_2+p)+I,\textstyle\frac{1}{2}(p_2+q_2+p)-I;I} 
\nonumber \\
&& \times \frac{\rme{(p_2,q_2)p_2+q_2-p;I}{\hat T^{\frac{1}{2}p}}{(p_2,q_2)p_2+q_2;\frac{1}{2}q_2}}{\sqrt{2I+1}}\nonumber\\
&& =\sum_{s(m_p)}\cg{\frac{1}{2}q_2}{\frac{1}{2}q_2-s}{\frac{1}{2}p}{m_p}{I}{I}
\sum_{\nu_1J(n_1)I_n}\rcg{(p_1,q_1)}{\nu_1;J}{(\lambda,\mu)}{n_1;I_n}{(p_2,q_2)}{p_2+q_2;\frac{1}{2}q_2}\nonumber \\
&&\times \sum_{m_\nu m_n}\cg{J}{m_\nu}{I_n}{m_n}{\frac{1}{2}q_2}{\frac{1}{2}q_2-s}
\sum_{j_a}\frac{p!}{(2j_a)!(p-2j_a)!}\sum_{m_a m_b}
\cg{j_a}{m_a}{\frac{1}{2}p-j_a}{m_b}{\frac{1}{2}p}{m_p} \nonumber \\
&&\quad\times \left[\hat T^{j_a}_{m_a}\ket{(p_1,q_1)\nu;J}\right]\left[\hat T^{\frac{1}{2}p-j_a}_{m_b}
\ket{(\lambda,\mu)n;I_n}\right]\, .
\eeqa
and one obtains
%%%%%
\beqa
&&\rcg{(p_1,q_1)}{\nu'_1;J'}{(\lambda,\mu)}{n'_1 I'_n}{(p_2,q_2)}{p_2+q_2-p;I}
\cg{J'}{m'_\nu}{I'_n}{m'_n}{I}{I}\nonumber \\
&&\qquad\times \frac{\rme{(p_2,q_2)p_2+q_2-p;I}{\hat T^{\frac{1}{2}p}}{(p_2,q_2)p_2+q_2;\frac{1}{2}q_2}}{\sqrt{2I+1}}\nonumber \\
&&\quad=\sum_{s(m_p)}\cg{\frac{1}{2}q_2}{\frac{1}{2}q_2-s}{\frac{1}{2}p}{m_p}{I}{I}
\sum_{\nu_1J(n_1)I_n}\rcg{(p_1,q_1)}{\nu_1;J}{(\lambda,\mu)}{n_1;I_n}{(p_2,q_2)}{p_2+q_2;\frac{1}{2}q_2}\nonumber \\
&&\times \sum_{m_\nu m_n}\cg{J}{m_\nu}{I_n}{m_n}{\frac{1}{2}q_2}{\frac{1}{2}q_2-s}
\sum_{j_a}\frac{p!}{(2j_a)!(p-2j_a)!}\sum_{m_a m_b}
\cg{j_a}{m_a}{\frac{1}{2}p-j_a}{m_b}{\frac{1}{2}p}{m_p} \nonumber \\
&&\times \bra{(p_1,q_1)\nu';J'}{\hat T^{j_a}_{m_a}}\ket{(p_1,q_1)\nu;J}
\bra{(\lambda,\mu)n';I'_n}{\hat T^{\frac{1}{2}p-j_a}_{m_b}}\ket{(\lambda,\mu)n;I_n}\, .
\eeqa 
Inserting the expressions for the matrix elements of the $\hat T^j$ operators produces an expression containing a 4-fold 
product of  $\sugroup{(2)}$ CGs, 
which can be reduced to 
an expression containing an $\sugroup{(2)}$ $9j$ symbol.
Technical manipulations then produces the final expression:
\beqa
\kern-5em &&\rcg{(p_1,q_1)}{\nu'_1;J'}{(\lambda,\mu)}{n'_1 ;I'_n}{(p_2,q_2)}{p_2+q_2-p;I}\nonumber \\
\kern-5em &&\quad =\frac{(-1)^{p-I-J'-I'_n}\sqrt{(2I+1)(q_2+1)(p+1)}}
{\rme{(p_2,q_2)p_2+q_2-p;I}{\hat T^{\frac{1}{2}p}}{(p_2,q_2)p_2+q_2;\frac{1}{2}q_2}}\nonumber \\
\kern-5em &&\times 
\sum_{\nu_1 (n_1)J I_n}
\rcg{(p_1,q_1)}{\nu_1 J}{(\lambda,\mu)}{n_1;I_n}{(p_2,q_2)}{p_2+q_2;\frac{1}{2}q_2}
\binom{p}{\nu_1-\nu'_1}(-1)^{\frac{1}{2}q_2+J+I_n} \nonumber \\
\kern-5em && \times 
\rme{(p_1,q_1)\nu'_1J'}{\hat T^{\frac{1}{2}(\nu_1-\nu'_1)}}{(p_1,q_1)\nu_1 J}
\rme{(\lambda,\mu)n'_1I'_n}{\hat T^{\frac{1}{2}(p-\nu_1+\nu'_1)}}{(\lambda,\mu)n_1 I_n}\nonumber \\
\kern-5em && \quad \times \left\{\begin{array}{ccc} 
I_n & J & \frac{1}{2}q_2 \\
\frac{1}{2}(p-\nu_1+\nu'_1)&\frac{1}{2}(\nu_1-\nu'_1) &\frac{1}{2}p \\
I'_n&J'&I\end{array}\right\} \label{eqn:generalexpression}
\eeqa
This expression has a clear interpretation.  From $\ket{(p_2,q_2)p_2+q_2,q_2-s,s;\frac{1}{2}q_2}$ 
we use Eq.(\ref{shiftedexpression}) to lower to the desired state in $(p_2,q_2)$.  It is then a matter of expanding Eq.(\ref{shiftedexpression}) using
$\hat C_{j1}=\hat C_{j1}^{(1)}+\hat C_{j1}^{(2)}$, with $\hat C_{j1}^{(i)}$ acting on states in $(p_1,q_1)$ for $i=1$ and on states
in  $(\lambda,\mu)$ for $i=2$.  The summation captures the expression of the highest weight state as a sum
of states in $(p_1,q_1)$ and $(\lambda,\mu)$.
The products of the type $(\hat C_{31}^{(i)})^{s_i-m_i}(\hat C_{21}^{(i)})^{s_i+m_i}$ are then re-expressed as tensors so
Eq.(\ref{eq:kpowertensor}) can be used to clean up the final result.  

\section{Weyl symmetries of the $SU(3)$ CGs}\label{sec:Weylsymmetries}

\hdg{The weight of a state is related to the occupation numbers by $(n_1-n_2,n_2-n_3)$.  The permutation group
acts on the occupational basis by permuting $n_1,n_2,n_3$ and produces a Weyl reflection of the original weight
$(n_1-n_2,n_2-n_3)$.  In this section we discuss the effect of the permutation group on the CG coefficients
through its permutation of  occupation numbers of states in $(p_1,q_1)$ and $(\lambda,\mu)$, and deduce the
corresponding symmetry of the CG coefficient.}

\subsection{The action of the permutation group}

In order to discuss Weyl symmetries of $SU(3)$ CGs, we need to make a distinction between the states  constructed
by \hdg{direct product of} $(p_1,q_1)$ and $(\lambda,\mu)$, i.e. the states obtained from the highest weight state of Eq.(\ref{generalhws}) 
- in this section these will be denoted by $\ket{(p_2,q_2)\nu I}_k$ with the subindex $k$ - from those 
of Eq.(\ref{eq:fullsu3states}), denoted in this section by $\ket{(p_2,q_2)\nu I}$ without the subindex $k$ .  
The difference is, up to an overall normalization, a power of the determinant factor of Eq.(\ref{eq:determinantfactor}).

The permutation group $S_3$ acts on the operators 
$\{\hat a_{1\alpha}^\dagger,\hat a_{2\alpha}^\dagger,\hat a_{3\alpha}^\dagger\}$ as
\beq
P_{\sigma}\hat a_{i\alpha}^\dagger P^{-1}_{\sigma}= \hat a_{\sigma(i)\alpha}^\dagger\, , \qquad \sigma\in S_3\, .
\eeq
Thus, consider now 
\beq
{\mbox{$\Bigl\langle {(p_1,q_1) \atop \nu ;J} ; {(\lambda,\mu) \atop n;I_n} \,\big\vert\, {(p_2,q_2) \atop N;I_N} \Bigr\rangle$}}_k
= {\mbox{$\Bigl\langle {(p_1,q_1) \atop \nu ;J} ; {(\lambda,\mu) \atop n ;I_n} \,\big\vert\, 
P^{-1}_{\sigma}P_{\sigma}\,\big\vert\,  {(p_2,q_2) \atop N ;I_N} \Bigr\rangle$}}_k \, . \label{eq:basicCGsymmetry}
\eeq
Because of the extra determinant factor in  $\ket{(p_2,q_2)N;I_N}_k$, the action of $P_\sigma$ on $\ket{(p_2,q_2)N;I_N}_k$ differs from the action 
of $P_\sigma$ on 
$\ket{(p_2,q_2)N;I_N}$ by a phase factor $(-1)^{k}$ when the permutation 
$\sigma$ contains an odd number of transpositions, as this corresponds to an odd number of transpositions of lines (or columns) in the determinant factor of  
Eq.(\ref{eq:determinantfactor}).    The determinant of course does not affect the absolute value of the CG since this determinant
transforms by the trivial irrep $(0,0)$.
Thus, defining
\beq
\sigma(k):= \left\{\begin{array}{ll}
k&\hbox{if } P_\sigma=P_{12}, P_{13}, P_{23}\, ,\\
0&\hbox{if }  P_{\sigma}=\unit, P_{123}, P_{132}
\end{array}\right.
\eeq 
we find $P_\sigma \ket{(p_2,q_2)N;I_N}_k= (-1)^{\sigma(k)} \ket{(p_2,q_2)N;I_N}$.  With this, we can expand Eq.(\ref{eq:basicCGsymmetry}) to obtain
\beqa
&&{\mbox{$\Bigl\langle {(p_1,q_1) \atop \nu;J} ; {(\lambda,\mu) \atop n;I_n} \,\big\vert\, {(p_2,q_2) \atop N;I_N} \Bigr\rangle$}}\nonumber \\
&&\, = (-1)^{\sigma(k)} 
\sum_{I'_NJ'I'_n} 
\bra{(p_1,q_1)\nu J} P^{-1}_{\sigma}\ket{(p_1,q_1)\nu' J'}
 \nonumber \\
&& \times \bra{(\lambda,\mu)n I_n} P^{-1}_{\sigma}\ket{(p_1,q_1)n' I'_n} \bra{(p_2,q_2)N' J'} P_{\sigma}\ket{(p_2,q_2)N J} \nonumber \\
&&\times
\cg{(p_1,q_1)}{\nu'; I'}{(\lambda,\mu)}{n';I'_n}{(p_2,q_2)}{N';I'_N}\, .
\eeqa 
All the states on the right hand side are now defined as per Eq.(\ref{eq:fullsu3states}), so the results of \cite{rowe1999representations} on the action
of $P_\sigma$ can be imported without modification.

\begin{figure}[h!]
\centering
\includegraphics[scale=0.5]{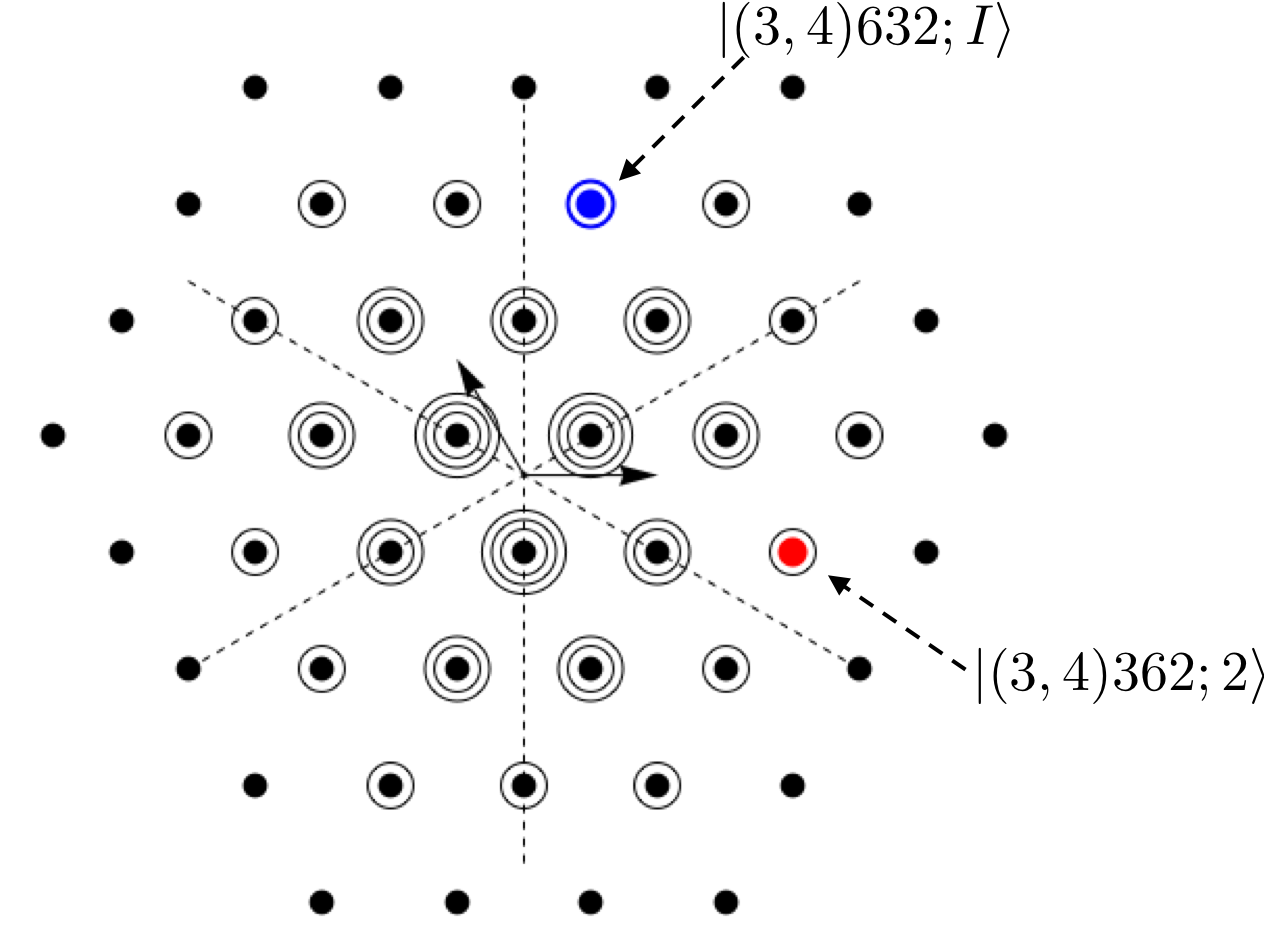}
\caption{The reflection of $\ket{(3,4){362;2}}$ is a linear combination of $\ket{(3,4)632;\frac{3}{2}}$ and  $\ket{(3,4)632;\frac{5}{2}}$.}
\label{fig:34CGreflection}
\end{figure}

If, for instance $(p_1,q_1),(\lambda,\mu),(p_2,q_2)$ are given respectively by $(5,1)$, $(4,0)$ and $(3,4)$, we have $k=0$ and 
easily find that the action of $P_{12}$ gives
\beqa
&&\cg{(5,1)}{241;\frac{3}{2}}{(4,0)}{121;\frac{3}{2}}{(3,4)}{362;2}
=\frac{2}{\sqrt{5}}\sqrt{\frac{7}{15}}
\cg{(5,1)}{421;\frac{3}{2}}{(4,0)}{211;1}{(3,4)}{632;\frac{5}{2}}\nonumber \\
&&\, +\frac{1}{\sqrt{5}}\left[
\sqrt{\frac{7}{15}} \cg{(5,1)}{421;\frac{3}{2}}{(4,0)}{211;1}{(3,4)}{632;\frac{3}{2}}
+\sqrt{\frac{8}{15}}
\cg{(5,1)}{421;\frac{1}{2}}{(4,0)}{211;1}{(3,4)}{632;\frac{3}{2}}\right], \\
&&\qquad\qquad =-\frac{7}{20\sqrt{2}} \, .
\eeqa 
A graphical illustration is given in Fig.\ref{fig:34CGreflection}.  The initial state of the irrep 
$\ket{(3,4){362;2}}$ is reflected to a linear combination of the $\ket{(3,4)632;\frac{3}{2}}$ and  $\ket{(3,4)632;\frac{5}{2}}$: the CGs involving
$\ket{(3,4){362;2}}$ are then linear combinations of those for the $\ket{(3,4)632;\frac{3}{2}}$ and  $\ket{(3,4)632;\frac{5}{2}}$ states.

This makes it clear that, in principle, we only need to obtain CGs for those states $\ket{(p_2,q_2)NJ}$ in the dominant Weyl sector.  
Those states have non-negative weights, so that $N_1\ge N_2\ge N_3$.  
The weight diagram of $\sugroup{(3)}$ contains $3!=6$ Weyl chambers, so in principle, one can ``save'' 
some computational effort in evaluating the CGs \cite{everdingcalculation}. 
Unfortunately, as the example above and the two examples
below illustrate, the Weyl relations usually involves expressing a generic CG as a linear combination
of CGs in the dominant Weyl sector, resulting in much less than a $6$-fold savings in the actual evaluation.
In addition, the Weyl symmetries are symmetries of the full rather than the reduced coefficients, meaning
that additional manipulations beyond the evaluation of reduced CGs are required to efficiently use these
symmetries.

\subsection{The example of $(1,1)\otimes (\sigma, \sigma)$}

The decomposition of the \hdg{direct product} $(1,1)\otimes(\sigma,\sigma)$ contains two copies of $(\sigma,\sigma)$,
which we label by $\rho=1,2$.  In such cases, the recursion 
relation of CG ends where CGs are expressed as combinations of two CGs involving the states of the $(1,1)$ irrep.  

We note that generators are components of an $\mathfrak{su}(3)$ tensor operators ${\cal T}^{(1,1)}$ and so, using the Wigner-Eckart theorem, their matrix elements contains Clebsch-Gordan coefficients.  Following \cite{hecht1965su3}\cite{draayer1973wigner}, we define the $\rho=1$ coefficients
so the matrix elements of the generators contain only $\rho=1$ CGs.  Coefficients with $\rho=2$ are obtained by orthogonality.
The table below gives explicit expressions for the highest weight state of each copy.

\begin{table}[h!] \centering   
{\renewcommand{\arraystretch}{2.25}
\begin{tabular}{| R | C |C | C | } \hline
&{\bigket{(\sigma,\sigma)}{2\sigma; \frac{1}{2}\sigma}}_1
&{\bigket{(\sigma,\sigma)}{2\sigma; \frac{1}{2}\sigma}}_2\\
\hline 
%%%%
\bigbra{(1,1)}{1;1}{(\sigma,\sigma)}{2\sigma,\frac{1}{2}\sigma}& 
%\sqrt{\frac{\sigma}{2(\sigma+1)(2\sigma+3)}} &\frac{1}{\sqrt{2(\sigma+1)}}&
\frac{1}{2}&-\frac{\sqrt{3}}{2}\sqrt{\frac{2\sigma+1}{2\sigma+3}}\\
\hline 
%%%%%
\bigbra{(1,1)}{1;0}{(\sigma,\sigma)}{2\sigma,\frac{1}{2}\sigma}
%&\sqrt{\frac{3(\sigma+2)}{2(\sigma+1)(2\sigma+3)}} &-\sqrt{\frac{3\sigma}{2(\sigma+1)(\sigma+2)}}
&\frac{\sqrt{3}}{2}\sqrt{\frac{\sigma}{\sigma+2}}
&\frac{1}{2}\sqrt{\frac{\sigma(2\sigma+1}{(\sigma+2)(2\sigma+3)}}\\
\hline 
%%%%%
\bigbra{(1,1)}{2;\frac{1}{2} }{(\sigma,\sigma)}{2\sigma-1,\frac{1}{2}(\sigma+1)}&
%-\frac{1}{\sigma+1}\sqrt{\frac{\sigma}{(2\sigma+3)}} &
%\frac{\sigma}{\sigma+1}&
\sqrt{\frac{\sigma+2}{2(\sigma+1)(\sigma+2)}}&
\sqrt{\frac{3(2\sigma+1)}{2(\sigma+1)(2\sigma+3)}}\\
\hline 
\bigbra{(1,1)}{2;\frac{1}{2}}{(\sigma,\sigma)}{2\sigma-1,\frac{1}{2}(\sigma-1)}&
%\frac{1}{\sigma+1}\sqrt{\frac{\sigma(2\sigma+1)(\sigma+2)}{(2\sigma+3)} }&
%\frac{1}{\sigma+1}\sqrt{\frac{2\sigma+1}{\sigma+2}}&
-\sqrt{\frac{(2\sigma+1)}{2(\sigma+1)(\sigma+2)}}&
\sqrt{\frac{3}{2(\sigma+1)(\sigma+2)(2\sigma+3)}} \\ \hline
\end{tabular}
\caption{Reduced CGs for the highest weight states of the two copies of $(\sigma,\sigma)$ in
$(1,1)\otimes(\sigma,\sigma).$}
}
\end{table}

As an application of the relations between CGs obtained from Weyl symmetries, we choose $\sigma = 3$ we start by writing the full Clebsch-Gordan coefficient in terms of the reduced CG and an SU(2) CG.  Thus for instance
\beq
\cg{(1,1)}{111;1}{(3,3)}{342;2}{(3,3)}{342;3}_\rho = \rcg{(1,1)}{1;1}{(3,3)}{3;2}{(3,3)}{3;3}_\rho \cg{1}{0}{2}{1}{3}{1}
\eeq 
Choosing for instance the permutation $P_{12}$ (though any permutation will do) and $\rho =2$ the full CG can be written 
\beqa 
&&\cg{(1,1)}{111;1}{(3,3)}{342;2}{(3,3)}{342;3}_2 = -\sum_{I_1,I_2,I_3}\rcg{(1,1)}{1;I_1}{(3,3)}{4;I_2}{(3,3)}{4;I_3}_2 \cg{I_1}{0}{I_2}{\frac{1}{2}}{I_3}{\frac{1}{2}}\nonumber\\
&&\times\langle (1,1)111;I_1|P_{12}|(1,1)111;1\rangle
\langle (3,3)432;I_2|P_{12}|(3,3)342;2\rangle \nonumber \\
&&\times \langle (3,3)432;I_3|P_{12}|(3,3)342;3\rangle \label{permutation1}
\eeqa
where the action of $P_{12}$ is given by
\beqa
&&\langle(\lambda,\mu)\nu_2\nu_1\nu_3;I'|P_{12}|(\lambda,\mu)\nu_1\nu_2\nu_3;I\rangle \nonumber \\
&&\kern-5em  = (-1)^{(\nu_3-2I-2I'+2\mu-\lambda)/2}\sqrt{(2I+1)(2I'+1)}  
\sixj{\nu_1/2}{\nu_3/2}{I'}{\nu_2/2}{\lambda/2}{I}
\eeqa
The possible values of $I_1$ are 0 and 1, $I_2 \hbox{ and } I_3$ can take the same values of $\myfrac{1}{2}, 
\myfrac{3}{2}$, or $\myfrac{5}{2}$:

\renewcommand{\arraystretch}{1.75}
\begin{center}
$\begin{array}{|l|c|c| c | } \hline 
&P_{12}|(1,1)111;1\rangle & P_{12}|(3,3)342;2\rangle &P_{12}|(3,3)342;3\rangle \\  \hline
\langle (1,1)111;0\vert &\frac{\sqrt{3}}{2}&&\\ \hline
\langle (1,1)111;1\vert  &-\frac{1}{2}&&\\ \hline
\langle (3,3)432;\textstyle\frac{1}{2}\vert &&-\frac{1}{2}& -\sqrt{\frac{7}{10}} \\ \hline
\langle (3,3)432;\textstyle\frac{3}{2}\vert && -\sqrt{\frac{2}{5}}& \frac{\sqrt{7}}{5}\\ \hline
\langle (3,3)432;\textstyle\frac{5}{2}\vert &&\frac{1}{2}\sqrt{\frac{7}{5}}& \frac{1}{5\sqrt{2}}\\ \hline
\end{array}$
\captionof{table}{All possible matrix elements for Eq.(\ref{permutation1}).  All empty entries are $0$.}
\end{center}

There are ten possible non-trivial combinations of $(I_1,I_2,I_3)$, thus the sum contains ten terms. 
Combining all of these terms, the sum gives an answer of 
\begin{equation}
\cg{(1,1)}{111;1}{(3,3)}{342;2}{(3,3)}{342;3}_2 = -\frac{2\sqrt{2}}{15}\, ,
\end{equation}
which (of course) is correct.

\section{Simpler cases of highest weight states}

\hdg{In this section we discuss some practical applications and specialized results.  In particular we discuss
the application of the recursion relation for highest weight 
of $(p_2,q_2)$ in $(p_1,q_1)\otimes (\lambda,0)$ with emphasis on the $(p_1,q_1)\otimes(2,0)$.  The resulting
table agrees with previous results given in \cite{hecht1965su3}.}

\subsection{Highest weight states for $(p_2,q_2)$ with $k=0$.}

\hdg{In this subsection (and this subsection only) we denote by $\mathfrak{su}_{ij}(2)$ the subalgebra of $\mathfrak{su}(3)$
spanned by $\{\hat C_{ij},\hat C_{ji}, [\hat C_{ij},\hat C_{ji}]\}$.  We denote the irreps of this subalgebra by $I_{ij}$.}

Suppose we have $(p_1,q_1)\otimes (\lambda,\mu)\to (p_2,q_2)$ so that
\begin{equation}
p_1+2q_1+\lambda+2\mu=p_2+2q_2\, .
\end{equation}
In other words,  we have a case where $k=0$ in Eq.(\ref{eq:kdefinition}).

As the occupation numbers then satisfy
$\nu_3+n_3=0$ for the highest weight state of $(p_2,q_2)$, the expansion of $\vert (p_2,q_2)\hbox{hw}\rangle$ is of the form
\beqa 
&&\vert (p_2,q_2)p_2+q_2,q_2,0;\textstyle\frac{1}{2}q_2\rangle  \nonumber \\
&& \qquad = \displaystyle\sum_{n_1n_2\nu_1\nu_2}
\cg{(p_1,q_1)}{\nu_1\nu_20;\frac{1}{2}\nu_2}
{(\lambda,\mu)}{n_1n_20;\frac{1}{2}n_2}{(p_2,q_2)}
{p_2+q_2,q_2,0;\frac{1}{2}q_2} \nonumber \\
&&\qquad\quad\,  \times \vert (p_1,q_1)\nu_1 \nu_2 0;\textstyle\frac{1}{2}\nu_2\rangle
\vert (\lambda,\mu)n_1n_2 0;\textstyle\frac{1}{2}n_2\rangle \label{hwstatek0case}
\eeqa  
Since the action of $\hat C_{k3}$ kills every state in Eq.(\ref{hwstatek0case}),
it is convenient instead to define a new basis where the states are coupled so as to form a multiplet of the subalgebra 
$\sugroup_{12}(2)$ spanned by $\hat C_{12},\hat C_{21}$ and $\hat C_{11}-\hat C_{22}$:
\beqa 
\rket{(\lambda,\mu)\nu_1\nu_2\nu_3;I_{12}}
&:=& \sum \cg{\frac{1}{2}\nu_1}{m_1}{\frac{1}{2}\nu_2}{m_2}
{I_{12}}{N_{12}}\cg{I_{12}}{N_{12}}{\frac{1}{2}\nu_3}{m_3}{\frac{1}{2}\lambda}{\frac{1}{2}\lambda}\nonumber \\
&&\qquad \qquad \times \ket{\textstyle\frac{1}{2}\nu_1,m_1}\ket{\textstyle\frac{1}{2}\nu_2,m_2}\ket{\textstyle\frac{1}{2}\nu_3,m_3}
\eeqa 
with the $\sugroup(2)_{12}\oplus\frak{u}(1)$ state denoted by a round ket for clarity.  
The original basis states are related to this new basis by 
\beqa 
&&( (\lambda,\mu) \nu_1\nu_2\nu_3;I_{12}\vert
(\lambda,\mu)\nu_1\nu_2\nu_3;\hdg{I_{23}}\rangle 
\nonumber \\
 &&\qquad 
= (-1)^{\lambda}\sqrt{(2\hdg{I_{23}}+1)(2I_{12}+1)}
\sixj{\textstyle\frac{1}{2}\nu_1}{\textstyle\frac{1}{2}\nu_2}{I_{12}}
{\textstyle\frac{1}{2}\nu_3}{\textstyle\frac{1}{2}\lambda}{\hdg{I_{23}}}\, ,
\eeqa 
where the subscripts $12$ and $23$ have been added to the angular momentum labels to identify the appropriate 
$\sugroup_{ij}(2)$ subalgebra. 
For $\nu_3=0$, this collapses to the simple expression
\begin{equation}
\ket{ (\lambda,\mu)\nu_1\nu_2 0;
\textstyle\frac{1}{2}\nu_2}
=(-1)^{\mu}\rket{(\lambda,\mu)\nu_1\nu_2 0;\textstyle\frac{1}{2}\lambda}\, .
\end{equation}
As could have been guessed, all the states $\vert (p_2,q_2),p_2+q_2,p_2,0;\frac{1}{2}q_2\rangle$ with $\nu_3=0$ are in fact, elements of a single 
$\sugroup_{12}(2)$ multiplet.  
The recursion relation generated by $\hat C_{12}$ will be, up to an overall phase, identical to the $\sugroup(2)$ recursion relation generated by $\hat L_+$ on $\sugroup(2)$ states.  As a result
the SU(3) CG for the highest $(p_2,q_2)$ weight state 
in the $\sugroup_{23}(2)$ basis is proportional to the overlap of states in the $\sugroup_{12}(2)$ basis:
\begin{equation}
\cg{(p_1,q_1)}{\nu_1\nu_20;\frac{1}{2}\nu_2}{(\lambda,\mu)}{n_1n_20; \frac{1}{2}n_2}{(p_2,q_2)}{p_2+q_2,q_2,0; \frac{1}{2}q_2}
=
\xi {\mbox{$\Bigl( {\frac{1}{2}p_1 \atop \frac{1}{2}(\nu_1-\nu_2)} ; {\frac{1}{2} \lambda \atop \frac{1}{2}(n_1-n_2)} \,\big\vert\, {\frac{1}{2} p_2 \atop \frac{1}{2} p_2} \Bigr)$}}
\end{equation}
with $\vert\xi\vert=1$.  If we now factor
\beqa 
&&\cg{(p_1,q_1)}{\nu_1\nu_20;\frac{1}{2}\nu_2}{(\lambda,\mu)}{n_1n_20; \frac{1}{2}n_2}{(p_2,q_2)}{p_2+q_2,q_2,0; \frac{1}{2}q_2}\nonumber \\
&&\quad =\,\rcg{(p_1,q_1)}{\nu_1;\frac{1}{2}(p_1-\nu_1)-q_2}{(\lambda,\mu)}{n_1;\frac{1}{2}(\lambda-n_1)-\mu}{(p_2,q_2)}{p_2+q_2;\frac{1}{2}q_2} 
\cg{\frac{1}{2}\nu_2}{\frac{1}{2}\nu_2}{\frac{1}{2}n_2}{\frac{1}{2}n_2}{\frac{1}{2}q_2}{\frac{1}{2}q_2}\, ,\\
&&\quad  =\,\rcg{(p_1,q_1)}{\nu_1;\frac{1}{2}(p_1-\nu_1)-q_1}{(\lambda,\mu)}{n_1;\frac{1}{2}(\lambda-n_1)-\mu}{(p_2,q_2)}{p_2+q_2;\frac{1}{2}q_2}
\eeqa 
since the regular SU(2) CG is $+1$, we find
\beqa
&&\rcg{(p_1,q_1)}{\nu_1;\frac{1}{2}(p_1-\nu_1)-q_1}{(\lambda,\mu)}{n_1;\frac{1}{2}(\lambda-n_1)-\mu}{(p_2,q_2)}{p_2+q_2;\frac{1}{2}q_2} \nonumber \\
&&\qquad =  \cg{ \frac{1}{2}p_1}{\nu_1-q_1-\frac{1}{2}p_1}{\frac{1}{2} \lambda}{n_1-\mu-\frac{1}{2}\lambda}{\frac{1}{2} p_2}{\frac{1}{2} p_2}
\eeqa
where the phase $\xi$ has been adjusted in accordance with the convention of Eq.(\ref{eq:phaseconvention}).

%%%%%%%%%%%%%%%%%%%%%%%%%%%%%%%%%%%%%%%%%%%%%%%%%%%%%%%%%%%%%%%%%
\subsection{Highest Weight state of $(p_2,q_2)$ in $(p_1,q_1)\otimes(\lambda,0)$.}

We look now at the recursion relation of Eqs.(\ref{recursion_reduced})  and (\ref{recursion_reduced2}) in the case 
$(p_1,q_1)\otimes(\lambda,0)$.  The decomposition of this direct product is multiplicity-free,  {\it i.e.}
an irrep $(p_2,q_2)$ occurs in this decomposition at most once 
\cite{speiser1964theory}\cite{wesslen2008geometric}\cite{o1982closed}.

%\subsubsection{The two recursions to construct the highest weight state}

In this special case, two key simplifications to the recursions for the highest weight state are that  
$I_n = I_n'+\frac{1}{2}$ and 
$I_n' = \frac{1}{2}(\lambda-n_1')$.  With these we can eliminate the sum 
in Eq.(\ref{recursion_reduced}), use exact expressions for the $6j$ symbol and matrix elements, 
and obtain an explicit relation between the steps in the recursion relation.
Those can further divided into two possible cases: one where $\tilde{J} = J'+\frac{1}{2}$ and one where $\tilde{J} = J'-\frac{1}{2}$. 

For $\tilde{J} = J'+\frac{1}{2}$, the recursion specializes to
\beqa
\kern-7em&&\rcg{(p_1,q_1)}{\nu_1-1;J'+\frac{1}{2}}{(\lambda,0)}{n_1';\frac{1}{2}(\lambda-n_1')}{(p_2,q_2)}{p_2+q_2;\frac{1}{2}q_2}\nonumber\\
\kern-7em&&= (-1)
\sqrt{\frac{2 n_1' (J'+1)(\frac{1}{2}(q_2+\lambda-n_1'+1)-J')(\frac{1}{2}(q_2-\lambda+n_1'+1)+J')}
{(2J'+1)(\frac{1}{2}(p_1+\nu_1)-J')(\frac{1}{2}(p_1-\nu_1)+J'+1)(\frac{1}{2}(p_1+2q_1-\nu_1)+J'+2)}}\nonumber\\
\kern-7em&&\qquad \times\rcg{(p_1,q_1)}{\nu_1;J'}{(\lambda,0)}{n_1'-1;\frac{1}{2}(\lambda-n_1'+1)}{(p_2,q_2)}{p_2+q_2;\frac{1}{2}q_2}\label{J_plus_recursion}
\eeqa
while for $\tilde{J} = J'-\frac{1}{2}$, the recursion relation becomes
\beqa
\kern-6em&&\rcg{(p_1,q_1)}{\nu_1-1;J'-\frac{1}{2}}{(\lambda,0)}{n_1';\frac{1}{2}(\lambda-n_1')}{(p_2,q_2)}{p_2+q_2;\frac{1}{2}q_2}\nonumber\\
\kern-6em&&= 
\sqrt{\frac{n_1'(2J')
(\frac{1}{2}(1-q_2+\lambda-n_1')+J')
(\frac{1}{2}(3+q_2+\lambda-n_1')+J')}
{(2J'+1)(\frac{1}{2}(p_1+\nu_1)+J'+1)(\frac{1}{2}(p_1+2q_1-\nu_1)-J'+1)(\frac{1}{2}(\nu_1-p_1)+J')}}\nonumber\\
\kern-5em&&\tab\times\rcg{(p_1,q_1)}{\nu_1;J'}{(\lambda,0)}{n_1'-1;\frac{1}{2}(\lambda-n_1'+1)}{(p_2,q_2)}{p_2+q_2;\frac{1}{2}q_2}\label{J_minus_recursion} 
\eeqa

%As an example of application, we give details of the first and second steps with the various combinations of  

\begin{figure}
    \centering
    \includegraphics[scale=0.5]{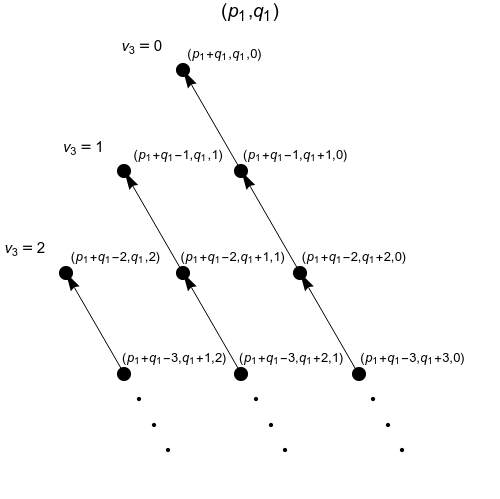}
    \caption{Various chains of $(p_1,q_1)$ states linked by the recursion of  Eq.(\ref{J_plus_recursion}).  States
    in each chain have constant $\nu_3$ occupation number.}
    \label{fig:p1q1states}
\end{figure}

We now make the following observation: if $\ket{(p_1,q_1)\bar{\nu_1}\bar{\nu_2}\bar{\nu_3};\hdg{I}}$ is 
a highest weight of the  
$\mathfrak{su}(2)$ subalgebra spanned by $\{\hat C_{23},\hat C_{32},\hat h_2\}$, 
so is $\hat C_{21}\ket{(p_1,q_1)\bar{\nu_1}\bar{\nu_2}\bar{\nu_3};\hdg{I}}$.  This follows by noting that an $\mathfrak{su}(2)$ highest weight state 
satisfies $\hat C_{32}\ket{(p_1,q_1)\bar{\nu_1}\bar{\nu_2}\bar{\nu_3};\hdg{I}}=0$ and that
\beq 
\hat C_{32}\hat C_{21}\ket{(p_1,q_1)\bar{\nu_1}\bar{\nu_2}\bar{\nu_3};\hdg{I }}=
\hat C_{21}\hat C_{32}\ket{(p_1,q_1)\bar{\nu_1}\bar{\nu_2}\bar{\nu_3};\hdg{I }}=0
\eeq 
by assumption on $\ket{(p_1,q_1)\bar{\nu_1}\bar{\nu_2}\bar{\nu_3};\hdg{I}}$.  Since
\beq 
\hat C_{21}\ket{(p_1,q_1)\bar{\nu_1}\bar{\nu_2}\bar{\nu_3};\hdg{I}}\propto 
\ket{(p_1,q_1)\bar{\nu_1}-1,\bar{\nu_2}+1,\bar{\nu_3};\hdg{I}+\textstyle\frac{1}{2}}\, ,
\eeq 
chains of highest weights with constant value of $\nu_3$ and with $\hdg{I}$ values increasing by $\frac{1}{2}$ can be used to generate a 
recursion for the reduced CGs of the appropriate multiplets.
The general situation is illustrated in Fig.\ref{fig:p1q1states}, where some of the $(p_1,q_1)$ occupation numbers needed to construct the highest weight of 
$(p_2,q_2)$ are identified, as well as arrows connecting states obtained using
the recursion of Eq.(\ref{J_plus_recursion}).  

The number of $\nu_3$-chains and their length is determined by $\lambda$ and $(p_2,q_2)$.   
We have $p_1+2q_1+\lambda=p_2+2q_2+3k$; the possible occupation numbers in
$(p_1,q_1)$ and $(\lambda,0)$ that can enter in the highest weight state of $(p_2,q_2)$ satisfy
\beq
\nu_1+n_1'=p_2+q_2+k\, ,\quad 
\nu_2+n_2'=q_2+k\, ,\quad 
\nu_3+n_3'=k\, .  \label{positivityrestriction}
\eeq 
Since $n'_i$ is non-negative, it is clear that the largest possible value of $\nu_3$ is $k$;  for $\nu_3=k$, the chain contains a single
state.

%%%%%%%%%%%%%%%%%%%%%%%%%%%%%%%%%%%%%%%%%%%%%%%%%%%

\subsubsection{The $\nu_3=0$ chain}

The states in $(p_1,q_1)$ with $\nu_3=0$ are in fact highest weight states of an 
$\sugroup{(2)}\oplus  \mathfrak{u}(1)$ multiplet
with $\hdg{I}=\frac{1}{2}\nu_2$, and can be used to obtain the reduced CGs for the multiplet.  Since 
$\nu_2$ increases by one as we go down the chain, the sequence of $\hdg{I}$ quantum numbers along this chain
is $\frac{1}{2}q_2, \frac{1}{2}(q_2+1), \frac{1}{2}(q_2+2)\ldots $ until $\frac{1}{2}(q_2+p_2)$.  

The $\nu_3=0$ chain consists of the sequence of $(p_1,q_1)$ states occupation numbers 
%states $(\nu_1\nu_2,\nu_3)$: 
$(p_1+q_1, q_1, 0)\gets (p_1+q_1-1, q_1+1, 0)\gets(p_1+q_1-2, q_1+2, 0)$.  
The state $\ket{(p_1,q_1)p_1+q_1,q_1,0;\frac{1}{2}q_1}$ is the highest weight state of $(p_1,q_1)$.
The first two steps of the recursion yield:
\beqa
&&\rcg{(p_1,q_1)}{p_1+q_1-1;\frac{1}{2}(q_1+1)}{(\lambda,0)}{n_1';\frac{1}{2}(\lambda-n_1')}{(p_2,q_2)}{p_2+q_2;\frac{1}{2}q_2}\nonumber\\
&&=-\frac{1}{2}\sqrt{\frac{n_1'(q_2-q_1+\lambda-n_1'+1)(q_2+q_1-\lambda+n_1'+1)}{p_1(q_1+1)}}\nonumber\\
&&\quad \times\rcg{(p_1,q_1)}{p_1+q_1;\frac{1}{2}q_1}{(\lambda,0)}{n_1'-1;\frac{1}{2}(\lambda-n_1'+1)}{(p_2,q_2)}{p_2+q_2;\frac{1}{2}q_2}\label{Jplusfirst}
\eeqa
and
\beqa
&&\rcg{(p_1,q_1)}{p_1+q_1-2;\frac{1}{2}(q_1+2)}{(\lambda,0)}{n_1'+1;\frac{1}{2}(\lambda-n_1'-1)}{(p_2,q_2)}{p_2+q_2;\frac{1}{2}q_2}\nonumber\\
&&=-\frac{1}{2}\sqrt{\frac{(n_1'+1)(q_2-q_1+\lambda-n_1'-1)(q_2+q_1-\lambda+n_1'+3)}{2(p_1-1)(q_1+2)}}\nonumber\\
&&\quad\times\rcg{(p_1,q_1)}{p_1+q_1-1;\frac{1}{2}(q_1+1)}{(\lambda,0)}{n_1';\frac{1}{2}(\lambda-n_1')}{(p_2,q_2)}{p_2+q_2;\frac{1}{2}q_2} \label{Jplussecond}
\eeqa
It is not hard to generalize the recursion to any step in the $\nu_3=0$ chain:
\beqa
\kern-3em&&\rcg{(p_1,q_1)}{p_1+q_1-(a+1);\frac{1}{2}(q_1+a+1)}{(\lambda,0)}{n_1'+a;\frac{1}{2}(\lambda-n_1'-a)}{(p_2,q_2)}{p_2+q_2;\frac{1}{2}q_2}\nonumber\\
\kern-3em&&=-\frac{1}{2}\sqrt{\frac{(n_1'+a)(q_2-q_1+\lambda-n_1'-2a+1)(q_2+q_1-\lambda+n_1'+2a+1)}{(a+1)(p_1-a)(q_1+a+1)}}\nonumber\\
\kern-3em&&\times\rcg{(p_1,q_1)}{p_1+q_1-a;\frac{1}{2}(q_1+a)}{(\lambda,0)}{n_1'+a-1;\frac{1}{2}(\lambda-n_1'-(a-1))}{(p_2,q_2)}{p_2+q_2;\frac{1}{2}q_2}
\eeqa
\medskip
%%%%%%%%%%%%%%%%%%%%%%%%%%%%%%%%%%%%%%5

\subsubsection{The $\nu_3=1$ chain}

The weights of
states with $\nu_3=1$ generally occurs twice: one of the state is in the same multiplet as a $\nu_3=0$ multiplet, so the
reduced CG for this case is known from the $\nu_3=0$ chain.  The remaining state with $\nu_3=1$ is the highest
weight state of an $\sugroup{(2)}\oplus  \mathfrak{u}(1)$ multiplet,
with $\hdg{I}=\frac{1}{2}(\nu_2-1)$ and can be used to compute the reduced CG for this multiplet, and so forth.

%It should be noted here that there is an important point to realize about these chains. 
While the $\nu_3=0$ chain exists for all $k$ values (recall $k=\frac{1}{3}(p_1+2q_1+\lambda-p_2-2q_2$)), 
the positivity restriction $\nu_3+n_3'=k$ of Eq.(\ref{positivityrestriction}) shows that, for the $\nu_3=1$ chain to exist, $k\ge 1$, and more generally
a chain with fixed $\nu_3$ will exist only if $k\ge \nu_3$. 
%one will never reach states in a $\nu_3$ chain which has a higher index than the k value one is working with (\ie for $k=0$, the $\nu_3 =1,2,3....$ states are not 
%attainable, for $k=1$ the $\nu_3 = 2,3,4....$ states are not attainable, and so on). This is because, if one tries to reach a state where $\nu_3 > k$, the quantum 
%numbers and angular momenta which the recursion relation desires will not be physically possible for at least one irrep.
%\medskip 

The first step of the $\nu_3=1$ chain, for instance,  
%here only consists of two states $(\nu_1,\nu_2,\nu_3)$: $(p_1+q_1-1,q_1,1)\Rightarrow(p_1+q_1-2,q_1+1,1)$, with the relation coming 
comes out out as
\beqa
&&\rcg{(p_1,q_1)}{p_1+q_1-2;\frac{1}{2}q_1}{(\lambda,0)}{n_1'+1;\frac{1}{2}(\lambda-n_1'-1)}{(p_2,q_2)}{p_2+q_2;\frac{1}{2}q_2}\nonumber\\
&&=-\frac{1}{2}\sqrt{\frac{(n_1'+1)(q_1+1)(q_2-q_1+\lambda-n_1'+1)(q_2+q_1-\lambda+n_1'+1)}{q_1p_1(q_1+2)}}\nonumber\\
&&\quad\times\rcg{(p_1,q_1)}{p_1+q_1-1;\frac{1}{2}(q_1-1)}{(\lambda,0)}{n_1';\frac{1}{2}(\lambda-n_1')}{(p_2,q_2)}{p_2+q_2;\frac{1}{2}q_2}
\eeqa
%%%%%%%%%%%%%%%%%%%%%%%%%%%%%%%%%%%%%%%%%%%%%%%%%%%
\subsection{Moving across $\nu_3$ chains} 

The recursion relation of Eq.(\ref{J_minus_recursion}) for $\tilde{J} = J'-\frac{1}{2}$ 
can be used to relate neighbouring chains of states in Fig.\ref{fig:p1q1states}, by connecting for instance   
$(p_1+q_1,q_1,0)\leftarrow (p_1+q_1-1,q_1,1)\leftarrow(p_1+q_1-2,q_1,2)$ and so forth if there are additional chains: 
\beqa
&&\rcg{(p_1,q_1)}{p_1+q_1-1;\frac{1}{2}(q_1-1)}{(\lambda,0)}{n_1';\frac{1}{2}(\lambda-n_1')}{(p_2,q_2)}{p_2+q_2;\frac{1}{2}q_2}\nonumber\\
&&=\frac{1}{2}\sqrt{\frac{n_1'(1-q_2+q_1+\lambda-n_1')(3+q_2+q_1+\lambda-n_1')}{(q_1+1)(p_1+q_1+1)}}\nonumber\\
&&\quad\times\rcg{(p_1,q_1)}{p_1+q_1;\frac{1}{2}{q_1}}{(\lambda,0)}{n_1'-1;\frac{1}{2}(\lambda-n_1'+1)}{(p_2,q_2)}{p_2+q_2;\frac{1}{2}q_2}\, ,
\label{eq:nu3across1} \\
&&\rcg{(p_1,q_1)}{p_1+q_1-2;\frac{1}{2}(q_1-2)}{(\lambda,0)}{n_1'+1;\frac{1}{2}(\lambda-n_1'-1)}{(p_2,q_2)}{p_2+q_2;\frac{1}{2}q_2}\nonumber\\
&&=\frac{1}{2}\sqrt{\frac{(n_1'+1)(-1-q_2+q_1+\lambda-n_1')(1+q_2+q_1+\lambda-n_1')}{2q_1(p_1+q_1)}}\nonumber\\
&&\quad \times\rcg{(p_1,q_1)}{p_1+q_1-1;\frac{1}{2}(q_1-1)}{(\lambda,0)}{n_1';\frac{1}{2}(\lambda-n_1')}{(p_2,q_2)}{p_2+q_2;\frac{1}{2}q_2}
\eeqa
\medskip

It is of course also possible to use Eq.(\ref{J_minus_recursion}) for $\tilde{J} = J'-\frac{1}{2}$ 
%The second chain which $\tilde{J} = J'-1/2$ 
to connect, for instance, $(p_1+q_1-1,q_1+1,0)\leftarrow (p_1+q_1-2,q_1+1,1)$:
\beqa
&&\rcg{(p_1,q_1)}{p_1+q_1-2;\frac{1}{2}q_1}{(\lambda,0)}{n_1'+1;\frac{1}{2}(\lambda-n_1'-1)}{(p_2,q_2)}{p_2+q_2;\frac{1}{2}q_2}\nonumber\\
&&=\frac{1}{2}\sqrt{\frac{(n_1'+1)(q_1+1)(1-q_2+q_1+\lambda-n_1')(3+q_2+q_1+\lambda-n_1')}{q_1(q_1+2)(p_1+q_1+1)}}\nonumber\\
&&\times\rcg{(p_1,q_1)}{p_1+q_1-1;\frac{1}{2}(q_1+1)}{(\lambda,0)}{n_1';\frac{1}{2}(\lambda-n_1')}{(p_2,q_2)}{p_2+q_2;\frac{1}{2}q_2} \, ,\label{Jminus_second}
\eeqa
but these reduced CGs are not new as they are accessible by moving to the top of the $\nu_3=1$ to the 
$\rcg{(p_1,q_1)}{p_1+q_1;\frac{1}{2}{q_1}}{(\lambda,0)}{n_1'-1;\frac{1}{2}(\lambda-n_1'+1)}{(p_2,q_2)}{p_2+q_2;\frac{1}{2}q_2}$ reduced CGs, and 
then going across to the $\nu_3=0$ chain using Eq.(\ref{eq:nu3across1}).

Systematic use of Eq.(\ref{J_minus_recursion}) for $\tilde{J} = J'-\frac{1}{2}$ will allow the expression of an CG at the top of any $\nu_3$ chain as
a multiple of the reduced CG of the $\nu_3=0$ chain of $(p_1,q_1)$.

\subsection{Example: the specific case of $(p_1,q_1)\otimes(2,0)$.}

\begin{table}\centering
\renewcommand{\arraystretch}{2.4}
\caption{Reduced CGs for highest weight of various $(p_2,q_2)$ irreps for the tensor product $(p_1,q_1)\otimes(2,0)$.}
\label{cg20case}
\begin{tabular}{| C | C | C | C | C | } 
\hline
k&(p_2,q_2)&(\nu_1 ;I_1) & ( n_1 ; I_2)&\rcg{(p_1,q_1)}{\nu_1 I_1}{(2,0)}{n_1 I_2}{(p_2,q_2)}{p_2+q_2 ; \frac{1}{2}q_2} \\ \hhline{|=|=|=|=l=|}
0&(p_1+2,q_1)& (p_1+q_1;\frac{1}{2}q_1)  & (2;0)& 1 \\ \hhline{|=|=|=|=|=|}
%%%%%%%%%%%%%%%%%%%%%%%%%%%%%%
0& (p_1,q_1+1)&(p_1+q_1;\frac{1}{2} {q_1}) & ( 1;\frac{1}{2}) & \sqrt{\displaystyle\frac{p_1}{p_1+2}}
\\ \hline
%%%%%%%%%%%%%%%%%%%%%%%%%%%%%%%%%%%
&&(p_1+q_1-1;\frac{1}{2}(q_1+1)) & (2;0) & -\sqrt{\displaystyle\frac{2}{p_1+2}}\\ \hhline{|=|=|=|=|=|}
%%%%%%%%%%%%%%%%%%%%%%%%%%%%%%%%%%%%
0&(p_1-2,q_1+2)&(p_1+q_1;\frac{1}{2}{q_1})& (0;1)& \sqrt{\displaystyle\frac{p_1-1}{p_1+1}}\\ \hline
%%%%%%%%%%%%%%%%%%%%%%%%%%%%%%%%%%%%
%(p_1,q_1)\otimes(2,0)\to (p_1-2,q_1+2)
&&(p_1+q_1-1;\frac{1}{2}(q_1+1)& (1;\frac{1}{2})&-\sqrt{\displaystyle\frac{2(p_1-1)}{p_1(p_1+1)}}\\ \hline
%%%%%%%%%%%%%%%%%%%%%%%%%%%%%%%%%%%%
%(p_1,q_1)\otimes(2,0)\to (p_1-2,q_1+2)
&&(p_1+q_1-2;\frac{1}{2}(q_1+2)) & (2,0)&\sqrt{\displaystyle\frac{2}{p_1(p_1+1)}}\\   \hhline{|=|=|=|=|=|}
%%%%%%%%%%%%%%%%%%%%%%%%%%%%%%%%%%%%%%%%%%%%%%%%%%%%%%%%%%%%%%%%%%%%%%%%%%%
%(p_1,q_1)\otimes(2,0)\to 
1&(p_1+1,q_1-1)&
(p_1+q_1;\frac{1}{2}{q_1}) & (1;\frac{1}{2})
%\rcg{(p_1,q_1)}{p_1+q_1;\frac{1}{2}{q_1}}{(2,0)}{1;\frac{1}{2}}{(p_1+1,q_1-1)}{p_1+q_1;\frac{1}{2}(q_1-1)}&=
&\sqrt{\displaystyle\frac{q_1+p_1+1}{q_1+p_1+3}}\\ \hline
%%%%%%%%%%%%%%%%%%%%%%%%%%%%%%%%%%%%
%(p_1,q_1)\otimes(2,0)\to (p_1+1,q_1-1)&\rcg{(p_1,q_1)}{
&&(p_1+q_1-1;\frac{1}{2}(q_1-1))&(2;0) %(p_1+1,q_1-1)}{p_1+q_1;\frac{1}{2}(q_1-1)}&=&\
&\sqrt{\displaystyle\frac{2}{q_1+p_1+3}}\\  \hhline{|=|=|=|=|=|}
%%%%%%%%%%%%%%%%%%%%%%%%%%%%%%%%%%%%
1& (p_1-1,q_1)%&\rcg{(p_1,q_1)}{p_1+q_1;\frac{1}{2}{q_1}}{(2,0)}{0;1}{(p_1-1,q_1)}{p_1+q_1-1;\frac{1}{2}{q_1}}
&(p_1+q_1;\frac{1}{2}{q_1})&(0;1)&\sqrt{\displaystyle\frac{p_1(q_1+p_1+1)}{(p_1+1)(q_1+p_1+2)}}\\  \hline
%%%%%%%%%%%%%%%%%%%%%%%%%%%%%%%%%%%%
&&(p_1+q_1-1;\frac{1}{2}(q_1+1))&(1;\frac{1}{2})&
-\sqrt{\displaystyle\frac{q_1(q_1+p_1+1)}{(1+q_1)(1+p_1)(q_1+p_1+2)}}\\ \hline
%%%%%%%%%%%%%%%%%%%%%%%%%%%%%%%%%%%%
&&(p_1+q_1-1;\frac{1}{2}(q_1-1)) & (1;\frac{1}{2})
&\sqrt{\displaystyle\frac{p_1(q_1+2)}{(1+q_1)(1+p_1)(q_1+p_1+2)}}\\ \hline
%%%%%%%%%%%%%%%%%%%%%%%%%%%%%%%%%%%%
&&(p_1+q_1-2;\frac{1}{2}{q_1})&(2;0)&-\sqrt{\displaystyle\frac{2}{(1+p_1)(q_1+p_1+2)}}\\ \hhline{|=|=|=|=|=|}
%%%%%%%%%%%%%%%%%%%%%%%%%%%%%%%%%%%%
2&(p_1,q_1-2)
&(p_1+q_1;\frac{1}{2}{q_1})&(0;1)&\sqrt{\displaystyle\frac{q_1+p_1}{q_1+p_1+2}}\\ \hline
%%%%%%%%%%%%%%%%%%%%%%%%%%%%%%%%%%%%
&&(p_1+q_1-1;\frac{1}{2}(q_1-1))&(1;\frac{1}{2})& %&=&
\sqrt{\displaystyle\frac{2(q_1+p_1)}{(q_1+p_1+1)(q_1+p_1+2)}}\\ \hline
%%%%%%%%%%%%%%%%%%%%%%%%%%%%%%%%%%%%
%(p_1,q_1)\otimes(2,0)\to (p_1,q_1-2)&\rcg{(p_1,q_1)}{p_1+q_1-2;\frac{1}{2}(q_1-2)}{(2,0)}{2;0}{(p_1,q_1-2)}{p_1+q_1-2;\frac{1}{2}(q_1-2)}&=
&&(p_1+q_1-2)&(2,0)&\sqrt{\displaystyle\frac{2}{(q_1+p_1+1)(q_1+p_1+2)}} \\ \hline
%%%%%%%%%%%%%%%%%%%%%%%%%%%%%%%%%%%%
\end{tabular}
\end{table}

Equations (\ref{Jplusfirst}-\ref{Jminus_second}) in the previous section are all we need to find explicit expressions for the case where $\lambda = 2$. The six possible $(p_2,q_2)$ irreps here are $(p_2,q_2) = \{(p_1+2,q_1), (p_1,q_1+1),
(p_1,q_1-2), (p_1-2,q_1+2), (p_1+1,q_1-1), (p_1-1,q_1)\}$. Using various substitutions to obtain each reduced 
CG in terms of the $(p_1,q_1)$ highest weight state, and knowing that the sum of the squares of reduced CGs must add to one, we can obtain
exact expressions for the separate reduced CGs as shown in Table \ref{cg20case}.  
\hdg{The expressions agree with, and expand Table 2 of 
\cite{hecht1965su3} for the construction of the highest weight state; 
indeed the results of this table for $(4,0)$ can also be used as the seed coefficients for the recursion relations if results 
for $(p_1,q_1)\otimes (4,0)$ are required.}

\section{Additional symmetries of $(p_1,q_1)\otimes(\lambda,0)\to (p_2,q_2)$}\label{sec:additional}

In addition to the general Weyl symmetries of Sec.\ref{sec:Weylsymmetries}, one can obtain further symmetries between the coupling
coefficients for 
$(p_1,q_1)\otimes(\lambda,0)\to (p_2,q_2)$ and those for $(\lambda,0)\otimes(p_1,q_1)\to (p_2,q_2)$ and 
$(p_2,q_2)\otimes(0,\lambda)\to (p_1,q_1)$.

The recursion relation for the  direct product $(p_1,q_1)\otimes(\lambda,0)\to (p_2,q_2)$ is identical to that for $(\lambda,0)\otimes(p_1,q_1)\to (p_2,q_2)$ but the first ``seed'' term in each recursion is different.  In particular, the seed coefficient
for $(\lambda,0)\otimes(p_1,q_1)\to (p_2,q_2)$ will differ from the corresponding reduced CG for the 
\hdg{product} $(p_1,q_1)\otimes(\lambda,0)\to (p_2,q_2)$ by at most a phase, which trickles through the rest of the evaluation
of the reduced CGs.  

Also, by taking the complex conjugate of the integral relation 
\beqa
&\int d\Omega D^{(p_1,q_1)}_{\nu' J';\hbox{\scriptsize hw}} (\Omega)D^{(\lambda,0)}_{n' I'_n;\bar n\bar I} (\Omega)
(D^{(p_2,q_2)}_{\mu I;\hbox{\scriptsize hw}}(\Omega) \nonumber \\
&= \frac{\hbox{vol}}{\hbox{dim}(p_2,q_2)} 
\cg{(p_1,q_1)}{\nu'J'}{(\lambda,0)}{n'I'_n}{(p_2,q_2)}{\mu I}
\cg{(p_1,q_1)}{\hbox{\scriptsize hw}}{(\lambda,0)}{\bar n\bar I}{(p_2,q_2)}{\hbox{\scriptsize hw}}
\eeqa 
we can relate coupling coefficients for $(p_1,q_1)\otimes (\lambda,0)\to (p_2,q_2)$ to those in
$(p_2,q_2)\otimes (0,\lambda)\to (p_1,q_1)$.

\subsection{Interchanging $(\lambda,0)$ and $(p_1,q_1)$}

We start by finding the phase of $\rcg{(p_1,q_1)}{\nu_1;\frac{1}{2}q_2}{(\lambda,0)}{\lambda;0}
{(p_2,q_2)}{p_2+q_2;\frac{1}{2}q_2}$.  Careful bookkeeping of the phases in the recursion relation  (\ref{recursion_reduced}) shows that this phase is given by 
\beq 
\varepsilon:=(-1)^{q_2-q_1+k}\, .
\eeq  
Since 
$\rcg{(\lambda,0)}{\lambda;0}{(p_1,q_1)}{\nu_1;\frac{1}{2}q_2}{(p_2,q_2)}{p_2+q_2;\frac{1}{2}q_2}$ must be positive in the 
\hdg{product}
$(\lambda,0)\otimes (p_1,q_1)\to (p_2,q_2)$, we have 
\beq 
\rcg{(\lambda,0)}{\lambda;0}{(p_1,q_1)}{\nu_1;\frac{1}{2}q_2}{(p_2,q_2)}{p_2+q_2;\frac{1}{2}q_2}
=\varepsilon\rcg{(p_1,q_1)}{\nu_1 \frac{1}{2}q_2}{(\lambda,0)}{\lambda;0}
{(p_2,q_2)}{p_2+q_2;\frac{1}{2}q_2}\, ,
\eeq 
and more generally for the $(p_2,q_2)$ highest weight
\beqa
&&\rcg{(\lambda,0)}{n'_1I_n'}{(p_1,q_1)}{\nu_1'J'}{(p_2,q_2)}{p_2+q_2;\frac{1}{2}q_2}\nonumber \\
&&\quad =\varepsilon\,(-1)^{J'+I'_n-\frac{1}{2}q_2}\rcg{(p_1,q_1)}{\nu'_1 J'}{(\lambda,0)}{n_1'I'_n}
{(p_2,q_2)}{p_2+q_2;\frac{1}{2}q_2}\, .  \label{eq:lambda0p1q1hw}
\eeqa 

Consider now the interchange of ordering of representations in Eq.(\ref{eqn:generalexpression}).  On the left hand side
we have $\rcg{(\lambda,0)}{n_1';I_n'}{(p_1,q_1)}{\nu_1';J'}{(p_2,q_2)}{p_2+q_2-p;I}$ while on the right we have
a sum containing $\rcg{(\lambda,0)}{n'_1I_n'}{(p_1,q_1)}{\nu_1'J'}{(p_2,q_2)}{p_2+q_2;\frac{1}{2}q_2}$ and the $9j$-symbol
${\scriptsize\renewcommand{\arraystretch}{1.25}\left\{\begin{array}{ccc} 
J&I_n  & \frac{1}{2}q_2 \\
\frac{1}{2}(\nu_1-\nu'_1) &\frac{1}{2}(p-\nu_1+\nu'_1)&\frac{1}{2}p \\
J'&I'_n&I\end{array}\right\}}$, where the first and second column in the symbol 
have been permuted w/r to the expression for 
$(p_1,q_1)\otimes (\lambda,0)\to (p_2,q_2)$.  Undoing this permutation yields a phase,
and using Eq.(\ref{eq:lambda0p1q1hw}) we obtain the
final result:
\beqa
&&\rcg{(\lambda,0)}{n_1';I_n'}{(p_1,q_1)}{\nu_1';J'}{(p_2,q_2)}{p_2+q_2-p;I} \nonumber \\
&&=
(-1)^{p+J'+I'_n+I-k+q_1}
\rcg{(p_1,q_1)}{\nu_1';J'}{(\lambda,0)}{n_1';I_n'}{(p_2,q_2)}{p_2+q_2-p;I}\, .
\eeqa
 
\subsection{Interchanging $(p_1,q_1)$ and $(p_2,q_2)$}

To understand the interchange of $(p_1,q_1)$ and $(p_2,q_2)$, start from Eq.(\ref{eq:integralform})
take the complex conjugate of this to obtain
\beqa
&&\int d\Omega \left(D^{(p_1,q_1)}_{\scriptsize\hbox{ hw};\scriptsize\hbox{ hw}}(\Omega)\right)^* 
\left(D^{(\lambda,0)}_{n' I'_n; n' I'_n}(\Omega)\right)^*
D^{(p_2,q_2)}_{\scriptsize\hbox{ hw};\scriptsize\hbox{ hw}}(\Omega) \nonumber \\
&&\quad =\int d\Omega \left(D^{(p_1,q_1)}_{\scriptsize\hbox{ hw};\scriptsize\hbox{ hw}}(\Omega)\right)^* 
D^{(0,\lambda)}_{\tilde{n} \tilde{I}_n;\tilde{n} \tilde{I}_n}
(\Omega)
D^{(p_2,q_2)}_{\scriptsize\hbox{ hw};\scriptsize\hbox{ hw}}(\Omega)\, ,\\
&&\quad =\frac{\hbox{vol}}{\hbox{dim}(p_1,q_1)}
\left\vert \cg{(p_2,q_2)}{\scriptsize\hbox{ hw}}{(0,\lambda))}{\tilde n,\tilde I_n}{(p_1,q_1)}{\scriptsize\hbox{ hw}}\right\vert^2
\eeqa
where $\tilde n=(\lambda- n'_1,\lambda- n'_2,\lambda- n'_3)$ and $\tilde I_n= I'_n$ and 
\begin{equation}
\left(D^{(\lambda,0)}_{ n' I'_n;  n'  I'_n}
(\Omega)\right)^*=D^{(0,\lambda)}_{\tilde{n} \tilde{I}_n;\tilde{n} \tilde{I}_n}(\Omega)
\end{equation} has been used.  This last property can be verified from the definition of the $(\lambda,0)$ and $(0,\lambda)$ 
$D$-functions given in
\cite{rowe1999representations}.   By convention $\cg{(p_1,q_1)}{\scriptsize\hbox{ hw}}{(\lambda,0)}{ n'I'_n}{(p_2,q_2)}{\scriptsize\hbox{ hw}}$ and $\cg{(p_2,q_2)}{\scriptsize\hbox{ hw}}{(0,\lambda))}{\tilde n,\tilde I_n}{(p_1,q_1)}{\scriptsize\hbox{ hw}}$ are positive so we find the intermediate result
\beq
\cg{(p_2,q_2)}{\scriptsize\hbox{ hw}}{(0,\lambda))}{\tilde n,\tilde I_n}{(p_1,q_1)}{\scriptsize\hbox{ hw}}=
\sqrt{\frac{\hbox{dim}(p_1,q_1)}{\hbox{dim}(p_2,q_2)}}
\cg{(p_1,q_1)}{\scriptsize\hbox{ hw}}{(\lambda,0)}{n'I'_n}{(p_2,q_2)}{\scriptsize\hbox{ hw}}\, .
\eeq

We can now take the complex conjugate of the more general form
\beq
\int d\Omega D^{(p_1,q_1)}_{\nu'J';{\scriptsize{\hbox{hw}}}}(\Omega)D^{(\lambda,0)}_{\bar n \bar I_n;n'I'_n}(\Omega)
\left(D^{(p_2,q_2)}_{\mu I;{\scriptsize{\hbox{hw}}}}(\Omega)
\right)^*
\eeq
and use 
\beq
\left(D^{(\lambda,0)}_{\bar n \bar I_n;n'I'_n}(\Omega)\right)^*
=(-1)^{\bar n_2+q_2-q_1+k}
D^{(0,\lambda)}_{\tilde{n} \bar I_n;\tilde{n}'I'_n}(\Omega)
\eeq
with $\tilde{n}'=(\lambda-n_1',\lambda-n_2',\lambda-n_3')$ to find the general result
\beqa
&&\cg{(p_2,q_2)}{\nu' J'}{(0,\lambda)}{\tilde{n}',I'_n}{(p_1,q_1)}{\nu' J'}\nonumber \\
&&=(-1)^{n'_2+q_2-q_1+k} \sqrt{\frac{\hbox{dim}(p_1,q_1)}{\hbox{dim}(p_2,q_2)}}
\cg{(p_1,q_1)}{\nu' J'}{(\lambda,0)}{n' I'_n}{(p_2,q_2)}{\mu I}\, .
\eeqa

\section{Discussion and conclusion}

In this work, we have shown how the use of an occupation basis to construct $\mathfrak{su}(3)$ basis states produces simplifications in the 
expressions of matrix elements, now expressed entirely in terms of quantities known from $\mathfrak{su}(2)$ technology.  

Like \cite{rowe1997algebraic}, our results are based on a recursion relation to quickly evaluate the highest weight state.  Once this is done, 
we can call on $\mathfrak{su}(2)$ 
technology to produce a final result.  An approach to disentangling multiple copies of an irrep in the decomposition of the tensor product has 
recently been proposed in \cite{campoamor2019external}\cite{campoamor2019some} and could be used if one must
deal with such multiplicities.

One advantage of the occupation basis is that the Weyl group acts naturally on basis states by permuting occupation numbers: 
this makes it straightforward to recover Weyl symmetries of the full $\mathfrak{su}(3)$ coefficients.  We have shown in 
connection with these Weyl symmetries that the essential difference with
the corresponding $\mathfrak{su}(2)$ results is the possible presence of weight multiplicities, 
which leads  to a sum over these multiplicities in the symmetry relations
between Weyl-reflected coefficients.  This decreases the naive $3!$ savings anticipated from the geometry of the weight diagram.
Finally, we have tabulated in Table \ref{cg20case} and detailed in \ref{sec:lls} some useful cases to illustrate general procedure.

HdG would like to thank Mr. Jeffrey Nielsen for early numerical work on this problem, and Dr. Hossein Tavakoli Dinani for his help with 
expressions arising from integration of SU(3) group functions.  The work of HdG is supported by NSERC.

\appendix

\section{Some special cases for $\rme{(\sigma,\sigma)2\sigma-p;J}{T^{\frac{1}{2}p}}{(\sigma,\sigma)2\sigma;\hf\sigma}$}\label{sec:specialcase}

%\subsection{$(\lambda,\mu)=(\sigma,\sigma)$}

For $(\lambda,\mu)=(\sigma,\sigma)$, the reduced matrix element for 
$\hat T^{\frac{1}{2}p}$ of Eq.(\ref{rmegeneral})  collapses to
\beqa
\kern-2em&&\rme{(\sigma,\sigma)2\sigma-p;J}{\hat T^{\frac{1}{2}p}}{(\sigma,\sigma)2\sigma-p;I}\nonumber \\
\kern-2em&&=(-1)^{\frac{1}{2}(p-2J+\sigma)}
\sqrt{\displaystyle\frac{(2J+1)(2\sigma+1)!\sigma!p!}{(\sigma-J+\frac{1}{2}(\sigma-p))!(\sigma+J+\frac{1}{2}(\sigma-p)+1)!}}\, .
\label{rmesigmasigma}
\eeqa

Using this and the known expressions for the appropriate $6j$ symbols we find:
\beqa
\kern-2em&&\rme{(\sigma,\sigma)2\sigma-(p+1);I-\textstyle\frac{1}{2}}{\hat T^{\frac{1}{2}}}{(\sigma,\sigma)2\sigma-p;I} \nonumber \\
\kern-2em&&=-\sqrt{\textstyle(\sigma+I+\frac{1}{2}(\sigma-p)+1)(\frac{1}{2}(\sigma-p)+I)(\frac{1}{2}(\sigma+p)-I+1)}\, , \\
\kern-2em&&\rme{(\sigma,\sigma)2\sigma-(p+1);I+\textstyle\frac{1}{2}}{\hat T^{\frac{1}{2}}}{(\sigma,\sigma)2\sigma-p;I}\nonumber \\
\kern-2em&&= \sqrt{\textstyle(\sigma-I+\frac{1}{2}(\sigma-p))(\frac{1}{2}(\sigma+p)+I+2)(I-\frac{1}{2}(\sigma-p)+1)}\, ,
\eeqa
and 
\beqa
\kern-2em&&\rme{(\sigma,\sigma)2\sigma-(p-1);I+\textstyle\frac{1}{2}}{\bar{T}^{\frac{1}{2}}}{(\sigma,\sigma)2\sigma-p;I} \nonumber \\
\kern-2em&& = -\sqrt{\textstyle(\sigma+I+\frac{1}{2}(\sigma-p)+2)(\frac{1}{2}(\sigma-p)+I+1)(\frac{1}{2}(\sigma+p)-I)}\, ,\\
\kern-2em&&\rme{(\sigma,\sigma)2\sigma-(p-1);I-\textstyle\frac{1}{2}}{\bar{T}^{\frac{1}{2}}}{(\sigma,\sigma)2\sigma-p;I} \nonumber \\
\kern-2em&& = -\sqrt{\textstyle(\sigma-I+\frac{1}{2}(\sigma-p)+1)(\frac{1}{2}(\sigma-p)+I+1)(I-\frac{1}{2}(\sigma-p))}\, .
\eeqa

\section{The CGs $(\lambda,0)\otimes(0,\lambda)\to (\sigma,\sigma)$}\label{sec:lls}

\subsection{The highest weight state in $(\sigma,\sigma)$}

We are interested in the CG for the \hdg{direct product} $(\lambda,0)\otimes(0,\lambda)\to (\sigma,\sigma)$.  In this case we have the
index $k=\lambda-\sigma$.  The recursion relation for the highest weight specializes
to 
\beqa 
&&\rcg{(\lambda,0)}{\lambda-s-1;\frac{1}{2}(s+1)}{(0,\lambda)}{s+\sigma+1;\frac{1}{2}(s+\sigma+1)}{(\sigma,\sigma)}{2\sigma;\frac{1}{2}\sigma} \nonumber \\
&&=(s+2)\frac{\rme{(0,\lambda)s+\sigma;\hf(s+\sigma)}{T^{\frac{1}{2}}}{(0,\lambda)s+\sigma+1;\hf(s+\sigma+1)}}
{\rme{(\lambda,0)\lambda-s-1;\hf(s+1)}{T^{\frac{1}{2}}}{(\lambda,0)\lambda-s;\hf s}} \nonumber \\
&&\times \rcg{(\lambda,0)}{\lambda-s;\frac{1}{2}s}{(0,\lambda)}{s+\sigma;\frac{1}{2}(s+\sigma)}{(\sigma,\sigma)}{2\sigma;\frac{1}{2}\sigma} 
\frac{1}{\sqrt{(s+2)(s+\sigma+1)}}\, ,
\eeqa 
where
\beq
\sixj{\hf}{\hf s}{\hf(s+1)}{\hf\sigma}{\hf(s+\sigma+1)}{\hf(s+\sigma)}= \frac{(-1)^{1-s-\sigma}}{\sqrt{(s+2)(s+\sigma+1)}}
\eeq
has been used.  Now:
\beqa
&&\rme{(0,\lambda)s+\sigma;\hf(s+\sigma)}{T^{\frac{1}{2}}}{(0,\lambda)s+\sigma+1;\hf(s+\sigma+1)} \nonumber \\
&&\qquad\qquad\qquad   = (-1)\sqrt{(s+\sigma+1)(\lambda-s-\sigma)(s+\sigma+2)}\, ,\\
&&\rme{(\lambda,0)\lambda-s-1;\hf(s+1)}{T^{\frac{1}{2}}}{(\lambda,0)\lambda-s;\hf s}\nonumber \\
&&\qquad\qquad\qquad \qquad \qquad   = \sqrt{(s+1)(s+2)(\lambda-s)}\, ,
\eeqa 
so the recursion can be brought to the form
\beqa
&&\rcg{(\lambda,0)}{\lambda-s-1;\frac{1}{2}(s+1)}{(0,\lambda)}{s+\sigma+1;\frac{1}{2}(s+\sigma+1)}{(\sigma,\sigma)}{2\sigma;\frac{1}{2}\sigma}
\nonumber \\
&&\quad = - \sqrt{\frac{(s+\sigma+2)(\lambda-s-\sigma)}{(\lambda-s)(s+1)}}
\rcg{(\lambda,0)}{\lambda-s;\frac{1}{2}s}{(0,\lambda)}{s+\sigma;\frac{1}{2}(s+\sigma)}{(\sigma,\sigma)}{2\sigma;\frac{1}{2}\sigma}\label{lambda_zero_recursion}
\eeqa

Upon iterating this relation, we see that Eq.(\ref{lambda_zero_recursion}) can be generalized as 
\beqa
\kern-5em&&\rcg{(\lambda,0)}{\lambda-s-a;\half(s+a)}{(0,\lambda)}{s+\sigma+a;\half(s+\sigma+a)}{(\sigma,\sigma)}{2\sigma;\half\sigma}\nonumber\\
\kern-5em&&= (-1)^a\prod_{b=0}^{a-1} \sqrt{\frac{(s+\sigma+2+b)(\lambda-s-\sigma-b)}{(\lambda-s-b)(s+1+b)}}
%\nonumber\\
%&\tab\times
\rcg{(\lambda,0)}{\lambda-s;\half s}{(0,\lambda)}{s+\sigma;\half(s+\sigma)}{(\sigma,\sigma)}{2\sigma;\half\sigma}\label{lambda_recursion}
\eeqa
Taking $s$ to be zero, we obtain a relation from between any state in the chain, and the highest weight state
\beqa
&&\rcg{(\lambda,0)}{\lambda-a;\half a}{(0,\lambda)}{\sigma+a;\half(\sigma+a)}{(\sigma,\sigma)}{2\sigma;\half\sigma}\nonumber\\
&&= (-1)^a\prod_{b=0}^{a-1} \sqrt{\frac{(\sigma+2+b)(\lambda-\sigma-b)}{(\lambda-b)(1+b)}} % \nonumber\\
%&\times
\rcg{(\lambda,0)}{\lambda;0}{(0,\lambda)}{\sigma;\half(\sigma)}{(\sigma,\sigma)}{2\sigma;\half\sigma}\label{lambda_HW_recursion}
\eeqa
Thus, knowing the reduced CG for the highest weight, one can obtain any step in the recursion simply through Eq.(\ref{lambda_recursion}). 
Expanding this product with various values of $a$, it can be seen that the relation can be written quite simply
in terms of factorials
\beqa
&&\rcg{(\lambda,0)}{\lambda-a;\half a}{(0,\lambda)}{\sigma+a;\half(\sigma+a)}{(\sigma,\sigma)}{2\sigma;\half\sigma}\nonumber\\
&&= (-1)^a\sqrt{\frac{(\sigma+a+1)!(\lambda-\sigma)!(\lambda-a)!}{(\sigma+1)!(\lambda-\sigma-a)!(\lambda)!(a)!}} %\nonumber\\
%&\tab\times
\rcg{(\lambda,0)}{\lambda;0}{(0,\lambda)}{\sigma;\half(\sigma)}{(\sigma,\sigma)}{2\sigma;\half\sigma}\label{lambda_HW_final_recursion}
\eeqa
It is nice to note here, as confirmation that this expression is correct, that for $a=0$ the coefficient equals 1 as it should. It is also valuable to notice the limit on the possible values of $a$ :  $a\le\lambda-\sigma$. Making use of this
limit,  we can iterate Eq.(\ref{lambda_HW_final_recursion}) explicitly to obtain the final expression
\beqa 
&&\rcg{(\lambda,0)}{\lambda;0}{(0,\lambda)}{\sigma;\half(\sigma)}{(\sigma,\sigma)}{2\sigma;\half\sigma} 
= \sqrt{\frac{\lambda!(2+2\sigma)!}{\sigma!(2+\lambda+\sigma)!}}
\eeqa
This can then be substituted into Eq.(\ref{lambda_HW_final_recursion}) to obtain an expression for any state
\beqa 
&&\rcg{(\lambda,0)}{\lambda-a;\half a}{(0,\lambda)}{\sigma+a;\half(\sigma+a)}{(\sigma,\sigma)}{2\sigma;\half\sigma} \nonumber \\
&&\qquad = (-1)^a\sqrt{\frac{(\sigma+a+1)!(\lambda-\sigma)!(\lambda-a)!(2+2\sigma)!}{(\sigma+1)!(\lambda-\sigma-a)!a!\sigma!(2+\lambda+\sigma)!}}\label{any_lambda_zero_state}
\eeqa 
\medskip

\subsection{The Complete Recursion Relation}

To complete we now look at Eq.(\ref{eqn:generalexpression}), which takes the specialized form
\beqa
\kern-7em&&\rcg{(\lambda,0)}{\nu'_1;\frac{1}{2}(\lambda-\nu_1')}{(0,\lambda)}{\lambda+\sigma-\nu_1'-p; \frac{1}{2}(\lambda+\sigma-\nu_1'-p)}{(\sigma,\sigma)}{2\sigma-p;I}
\nonumber \\
\kern-7em&&\quad =\frac{(-1)^{I+\lambda+\frac{1}{2}\sigma-\nu_1'+\frac{1}{2}p}\sqrt{(2I+1)(\sigma+1)(p+1)}}
{\rme{(\sigma,\sigma)2\sigma-p;I}{\hat T^{\frac{1}{2}p}}{(\sigma,\sigma)2\sigma;\frac{1}{2}\sigma}}\nonumber \\
\kern-7em&&\times 
\sum_{\nu_1 }
\rcg{(\lambda,0)}{\nu_1;\frac{1}{2}(\lambda-\nu_1)}{(0,\lambda)}{\lambda+\sigma-\nu_1;\frac{1}{2}(\lambda+\sigma-\nu_1)}{(\sigma,\sigma)}{2\sigma;\frac{1}{2}\sigma}
\binom{p}{\nu_1-\nu'_1}(-1)^{\lambda+\sigma-\nu_1} \nonumber \\
\kern-7em&& \times 
\rme{(\lambda,0)\nu'_1;\textstyle\frac{1}{2}(\lambda-\nu'_1)}{\hat T^{\frac{1}{2}(\nu_1-\nu'_1)}}{(\lambda,0)\nu_1; \textstyle\frac{1}{2}(\lambda-\nu_1)} \nonumber \\
\kern-7em&& \times 
\rme{(0,\lambda)\lambda+\sigma-\nu_1'-p;\textstyle\frac{1}{2}(\lambda+\sigma-\nu_1'-p)}{\hat T^{\frac{1}{2}(p-\nu_1+\nu'_1)}}
{(0,\lambda)\lambda+\sigma-\nu_1;\textstyle\frac{1}{2}(\lambda+\sigma-\nu_1)}\nonumber \\
\kern-7em&& \quad \times 
\left\{\begin{array}{ccc} 
\frac{1}{2}(\lambda+\sigma-\nu_1) & \frac{1}{2}(\lambda-\nu_1) & \frac{1}{2}\sigma \\
\frac{1}{2}(p-\nu_1+\nu'_1)&\frac{1}{2}(\nu_1-\nu'_1) &\frac{1}{2}p \\
\frac{1}{2}(\lambda+\sigma-p-\nu'_1)&\frac{1}{2}(\lambda-\nu'_1)&I\end{array}\right\}\label{complete_lambda_zero_recursion}
\eeqa

The three matrix elements in this equation can be expressed as follows
\beqa
\kern-5em&&\langle(\sigma,\sigma)2\sigma-p;I\Vert\hat{T}^{\frac{1}{2}p}\Vert (\sigma,\sigma)2\sigma;\textstyle\frac{1}{2}\sigma\rangle\nonumber\\
\kern-5em&&\quad =(-1)^{(p-2I+\sigma)/2}\sqrt{\frac{(2I+1)(2\sigma+1)!\sigma!p!}{(\sigma-I+\textstyle\frac{1}{2}(\sigma-p))!(\sigma+I+\frac{1}{2}(\sigma-p)+1)!}}\\
\kern-5em&&\langle(\lambda,0)\nu_1';\textstyle\frac{1}{2}(\lambda-\nu_1')\Vert \hat{T}^{\frac{1}{2}(\nu_1-\nu_1')}
\Vert (\lambda,0)\nu_1;\textstyle\frac{1}{2}(\lambda-\nu_1)\rangle % \nonumber\\
%&\tab
=\sqrt{\displaystyle\frac{\nu_1! (\lambda+1-\nu'_1)!}{\nu'_1!(\lambda-\nu_1)!}}\\
\kern-5em&&\langle(0,\lambda)\lambda+\sigma-\nu_1'-p;\textstyle\frac{1}{2}(\lambda+\sigma-\nu_1'-p)\Vert \hat{T}^{\frac{1}{2}(p-\nu_1+\nu_1')}
\Vert (0,\lambda)\lambda+\sigma-\nu_1;\textstyle\frac{1}{2}(\lambda+\sigma-\nu_1)\rangle\nonumber\\
\kern-5em&&\quad
=(-1)^{2\lambda+2\sigma-\nu_1-\nu_1'-p}\sqrt{\frac{(\lambda+\sigma-\nu_1+1)!(\nu_1'+p-\sigma)!}{(\lambda+\sigma-\nu_1'-p)!(\nu_1-\sigma)!}}
\eeqa
We can use Eq.(\ref{any_lambda_zero_state}) with the substitution $\nu_1 = \lambda-a$, and properties of the $9j$ coefficients to obtain 
an explicit form for Eq.(\ref{complete_lambda_zero_recursion}):
\beqa
\kern-7em&&\rcg{(\lambda,0)}{\nu_1';\frac{1}{2}(\lambda-\nu_1')}{(0,\lambda)}{\lambda+\sigma-\nu_1'-p;\frac{1}{2}(\lambda+\sigma-\nu_1'-p)}{(\sigma,\sigma)}{2\sigma-p;I}
\nonumber\\
\kern-7em&&=(-1)^\lambda\Big(\frac{(\lambda-\nu_1')!}{\sigma!}\Big)\textstyle\sqrt{\frac{(\sigma-I+\half(\sigma-p))!(\sigma+I+\half(\sigma-p)+1)!(\lambda-\sigma)!(2\sigma+2)(\nu_1'+p-\sigma)!(1+I+\half(p+\sigma))!}{(2+\lambda+\sigma)!(\nu_1')!(\lambda-\nu_1'-I+\half(\sigma-p))!(1+I+\lambda-\nu_1'+\half(\sigma-p))!(\half(\sigma+p)-I)!}}\nonumber\\
\kern-7em&&\times\sum_{\max[\sigma,\nu_1']}^{\min[\lambda,\nu_1'+p,\lambda-I+\half(\sigma+p)]}(-1)^{\nu_1}\Big(\frac{(\lambda-I-\nu_1+\half(\sigma+p))!\nu_1!}{(\nu_1-\nu_1')!(p-\nu_1+\nu_1')!(\nu_1-\sigma)!(\lambda-\nu_1)!}\Big)\nonumber\\
\kern-7em&&
\textstyle{_3}F_2(\nu_1-\lambda, I+\nu_1'-\lambda+\frac{1}{2}(p-\sigma),1+I+\frac{1}{2}(\sigma-p);\nu_1'-\lambda,I+\nu_1-\lambda-\frac{1}{2}(p+\sigma);1) \nonumber \\
\kern-7em&& \label{eq:74}
\eeqa
The bounds of the sum are determined by the factorials involved with $\nu_1$.

Eq.(74) enters in the construction of irreducible $\sugroup{(3)}$ tensor operators \cite{martins2019correspondence}
\cite{klimov2017generalized}.  The combinations
\begin{equation}
\hat{T}^{\lambda}_{(\sigma,\sigma)\nu I}=\sum_{\nu' I' n I}\cg{(\lambda,0)}{\nu' I'}{(0,\lambda)}{n I}{(\sigma,\sigma)}{\nu I} (-1)^{n_2}
\ket{(\lambda,0)\nu' I'}\bra{(\lambda,0) n I}
\end{equation}
are tensor operators transforming by irrep $(\sigma,\sigma)$ and acting on states in $(\lambda,0)$.  The extra factor $(-1)^{n_2}$ has been introduced to 
guarantee the correct sign of the commutation relation of these tensors with generators.  

\section{Performance and the $(p_1,q_1)\otimes(\lambda,0)$ CGs}

We have constructed a Mathematica\texttrademark\ code \cite{su3githubcode} to implement the evaluation of reduced CG coefficients for  the
decomposition of $(p_1,q_1)\otimes(\lambda,0)$ as a special case of Eq.(\ref{eqn:generalexpression}), with the double objective of testing the performance of 
the algorithm of Eq.(\ref{eqn:generalexpression}) and investigating the behaviour of reduced CG in the limit of large representations.  
There is indeed considerable literature on the asymptotics of $\sugroup{(2)}$ coefficients 
\cite{de1995asymptotic}\cite{de1998clebsch}\cite{sprung2009asymptotic}\cite{wigner2012group}\cite{brussaard1957classical}
\cite{ponzano1968semiclassical}\cite{reinsch1999asymptotics}\cite{rowe2010shifted} and, given that our algorithm relies heavily on $\sugroup{(2)}$ for which
asymptotics are known, we hoped to identify some obvious trends in the $\mathfrak{su}(3)$ case.  Unfortunately we have not been able to produce 
useful asymptotic formulae using the appropriate approximations for the various $\mathfrak{su}(2)$ factors.

Various results of computing all reduced CGs of the type
\begin{equation}
 \rcg{75,60}{47,22}{53,0}{27,13}{(p_2,q_2)}{N_1,35}\label{plotclebsch}
\end{equation}
where $(p_2,q_2)$ range over all possible irreps in the decomposition of the \hdg{direct product}
$(75,60)\otimes(53,0)$, and $N_1 = 47+27-k$ as presented as histograms below.   The decomposition contains $1485$ irreps, none of which 
are repeated. 

\begin{figure}[h!]
    \centering
    \includegraphics[scale=0.4]{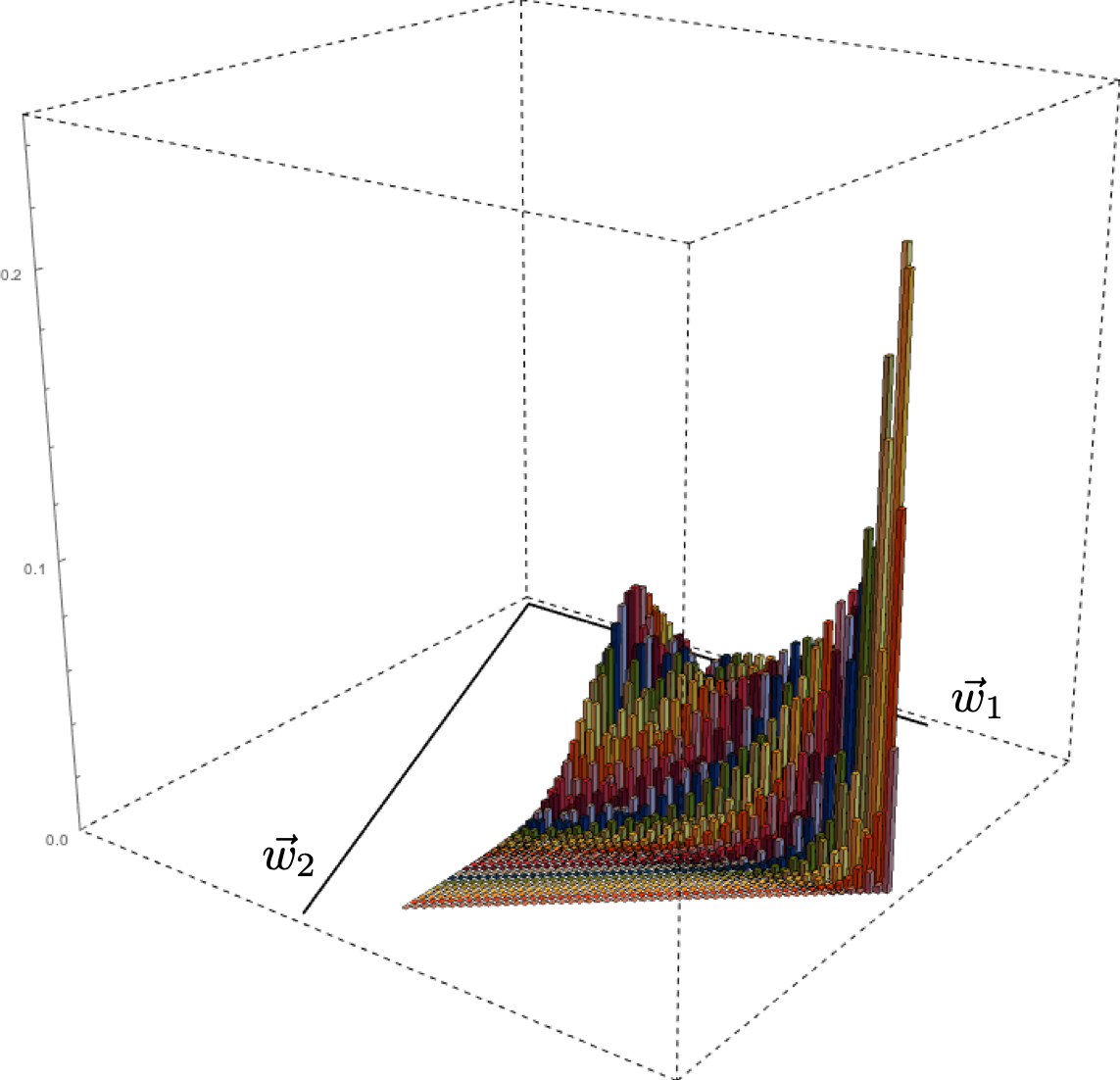} \\
    \includegraphics[scale=0.35]{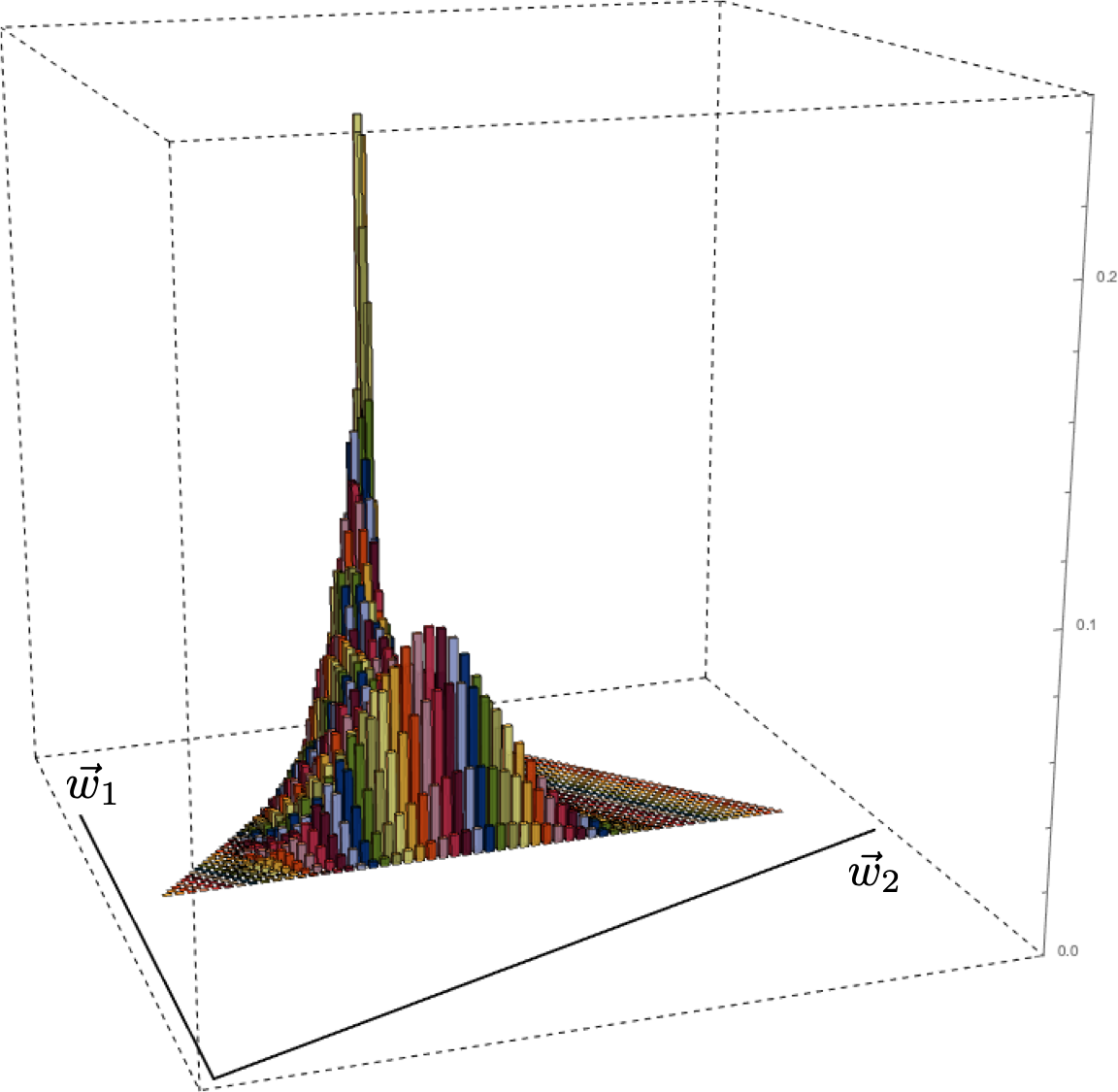}\\ 
    \includegraphics[scale=0.4]{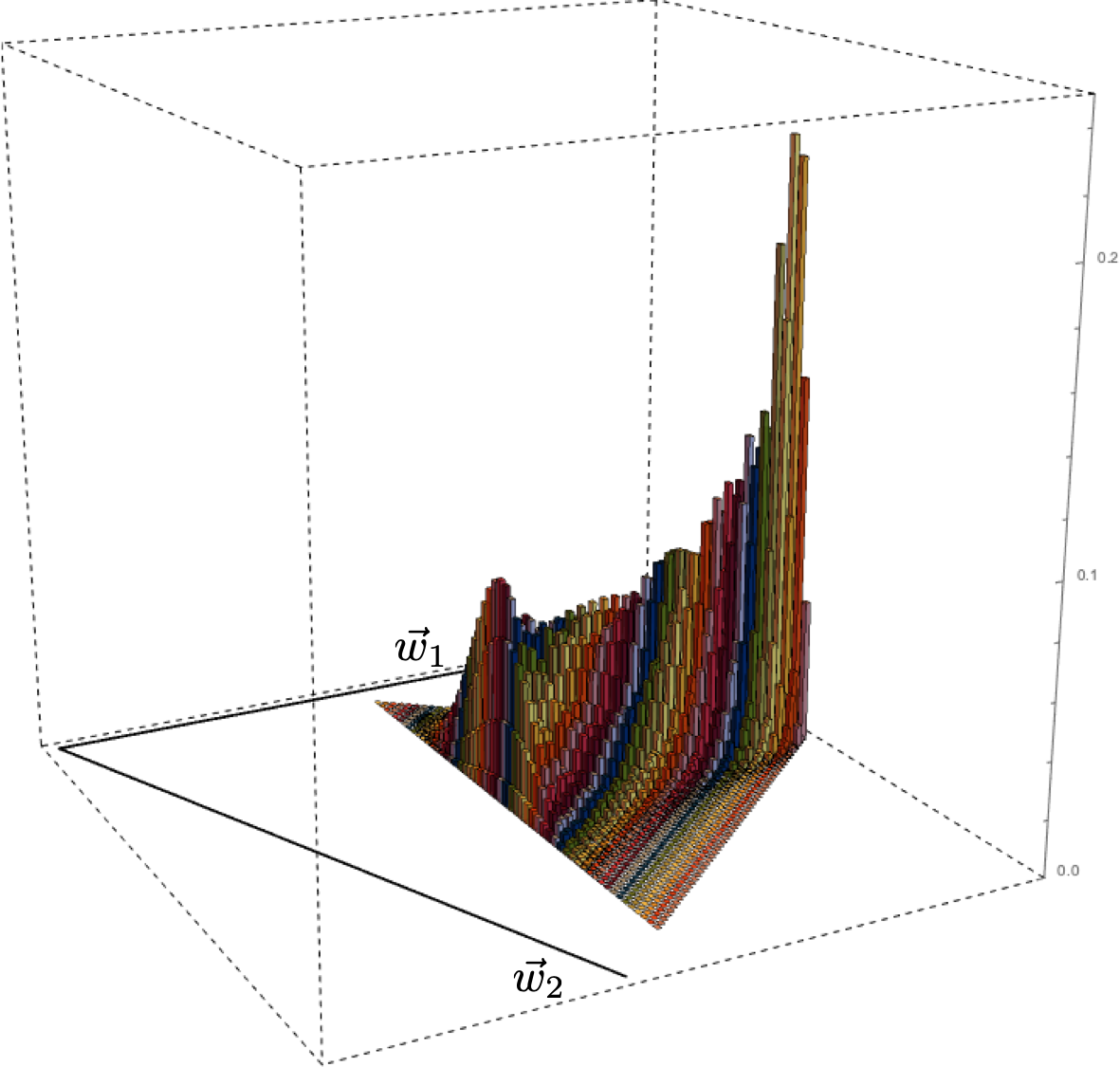}
    \caption{Three views of the absolute values of the 
    reduced CGs of Eq.(\ref{plotclebsch}).  CGs for $(p_2,q_2)$ irreps with the same value of $k$ have the same color and appear in a row.
    The directions along the fundamental weights $\vec w_1$ and $\vec w_2$ are also indicated; the intersection of these lines is the center
    of the weight diagram.}
    \label{fig:3dhistograms}
\end{figure}

\begin{figure}[h!]
    \centering
    \includegraphics[scale=0.4]{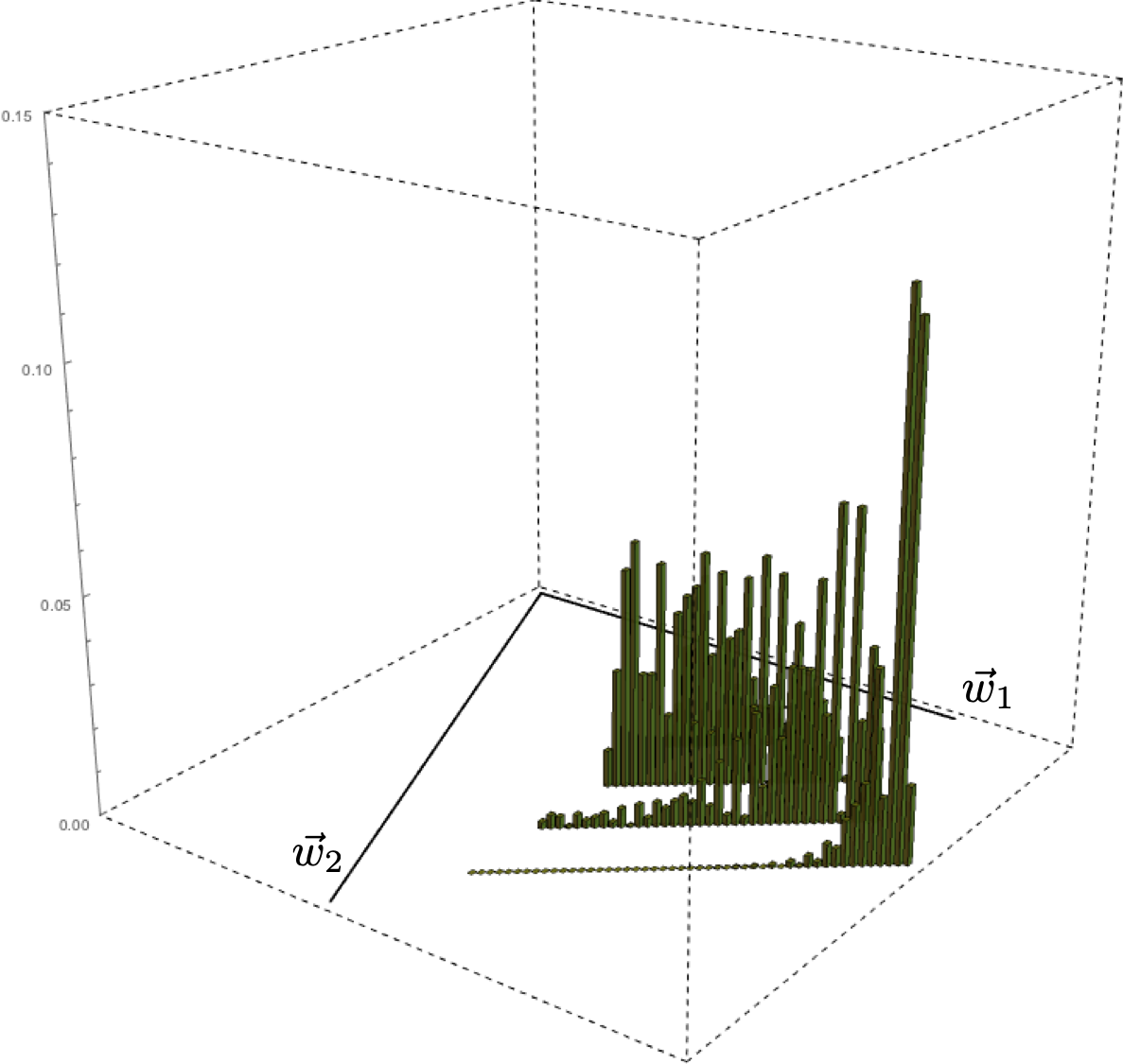} \\
    \includegraphics[scale=0.35]{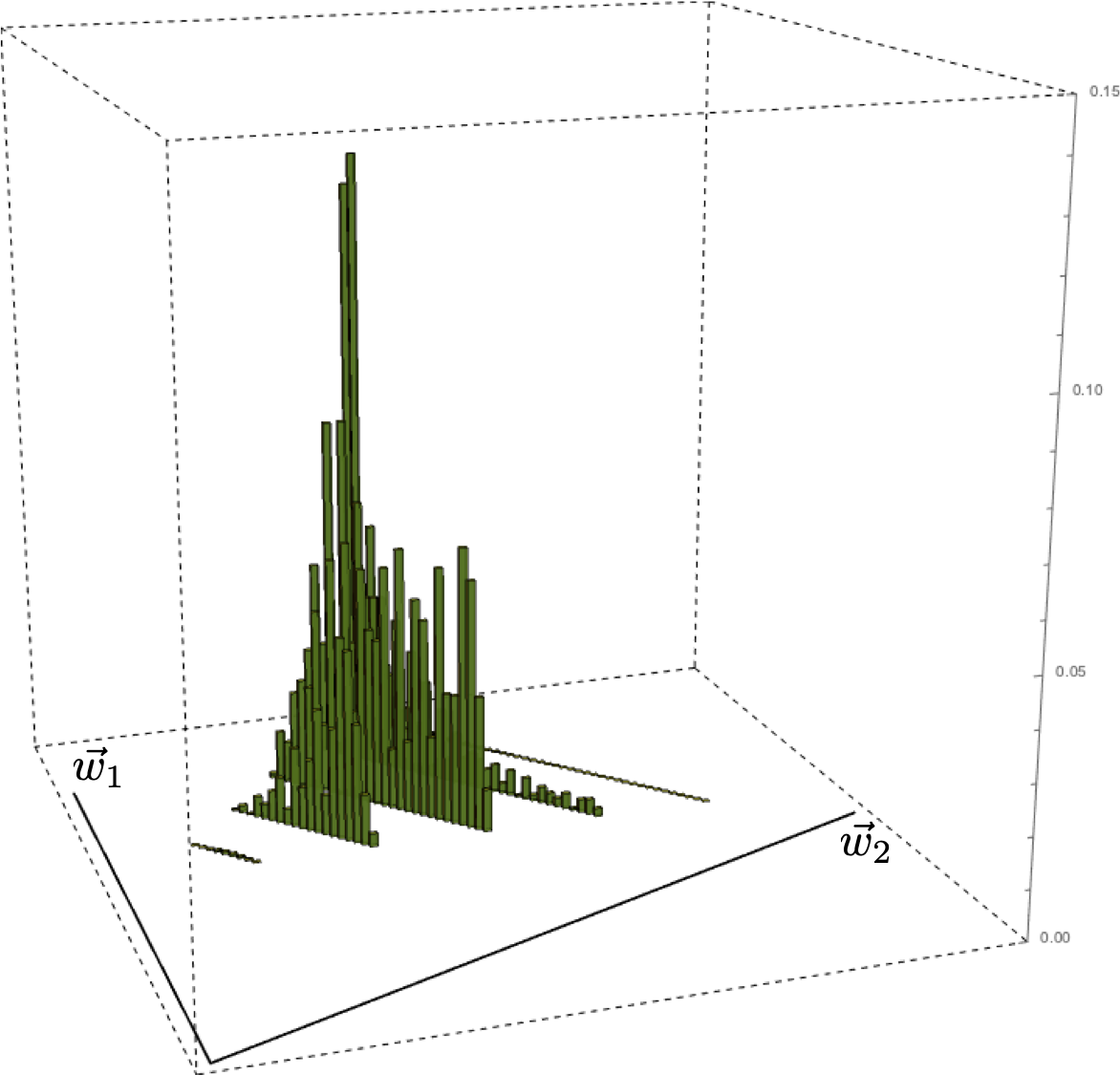}\\ 
    \includegraphics[scale=0.4]{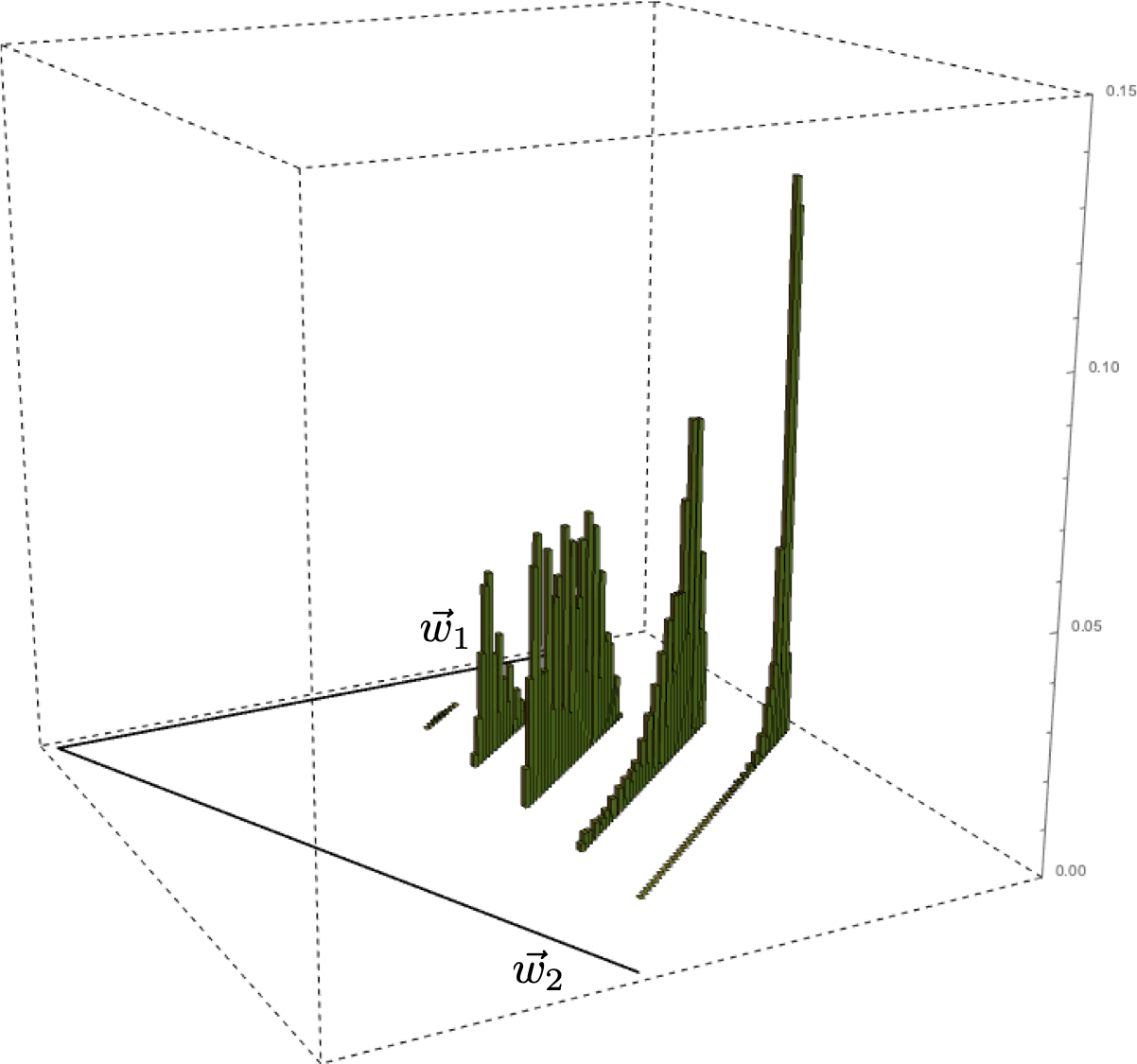}
    \caption{Three views of the absolute values of the 
    reduced CGs of Eq.(\ref{plotclebsch}) for values of $(p_2,q_2)$ with values $k=5,15,25,35$ and $45$.
     The directions along the fundamental weights $\vec w_1$ and $\vec w_2$ are also indicated; the intersection of these lines is the center
    of the weight diagram.}
    \label{fig:3dpartialhistograms}
\end{figure}

Fig. \ref{fig:3dhistograms} presents three views of the absolute values of the reduced CG of Eq.(\ref{plotclebsch}).  For contrast we have used different colors
for $(p_2,q_2)$ irreps with different values of $k$.  Because only 10 different colors are available, irreps with $k=0,10,20,30,40,50$ are of the same color, as are 
irreps with $k=1,11,21,31,41,51$. etc.  The values of $k$ range from $0$ to $53$.  
Fig. \ref{fig:3dpartialhistograms} shows three views of subsets of CGs of Fig.\ref{fig:3dhistograms} for irreps $(p_2,q_2)$ with
values of $k=5,15,25,35$ and $45$.

\begin{figure}[h!]
    \centering
    \includegraphics[scale=0.295]{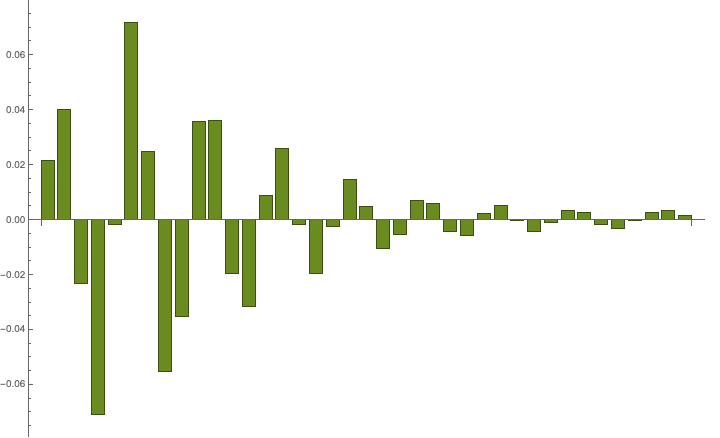} \includegraphics[scale=0.275]{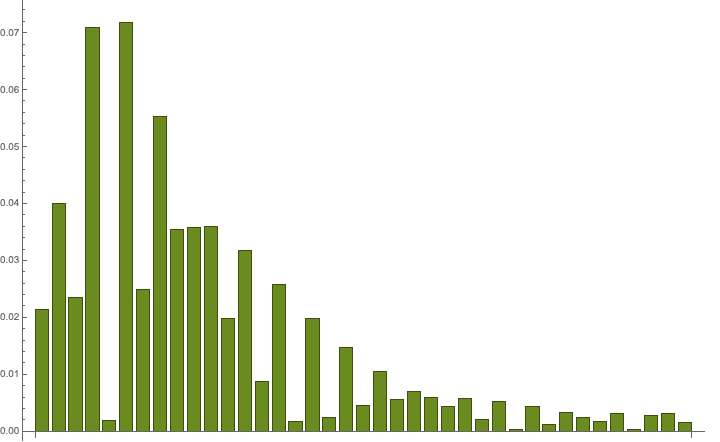}
    \caption{Histograms showing the exact and absolute values of Eq.(\ref{plotclebsch}) for $k=15$
    and $(p_2,q_2)$ values $(113,35), (111,46), (109,47)\ldots (37,83)$  }
    \label{fig:2dhistograms}
\end{figure}

In Fig.\ref{fig:2dhistograms}, we gives the actual (signed) reduced CG on the left, and the absolute value of this 
reduced CG on the right, for the 
reduced CG of Eq.(\ref{plotclebsch}) and $(p_2,q_2)$ irreps
with $k=15$.  The reduced CGs are given in left to right order for the $(p_2,q_2)$ values $(113,35), (111,46), (109,47)\ldots (37,83)$.

\begin{figure}[h!]
    \centering
    \includegraphics[scale=0.295]{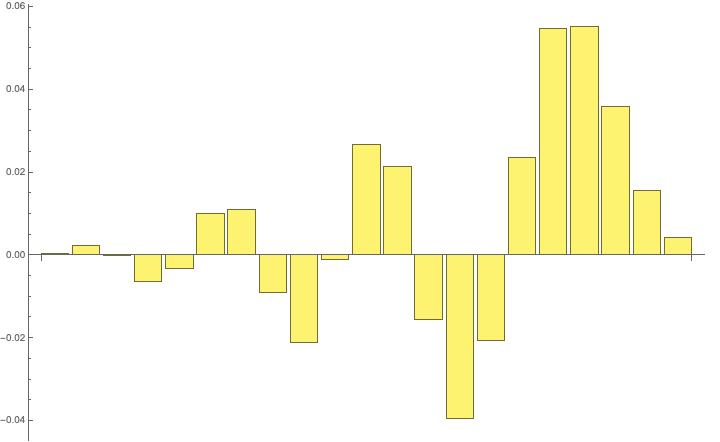} \includegraphics[scale=0.275]{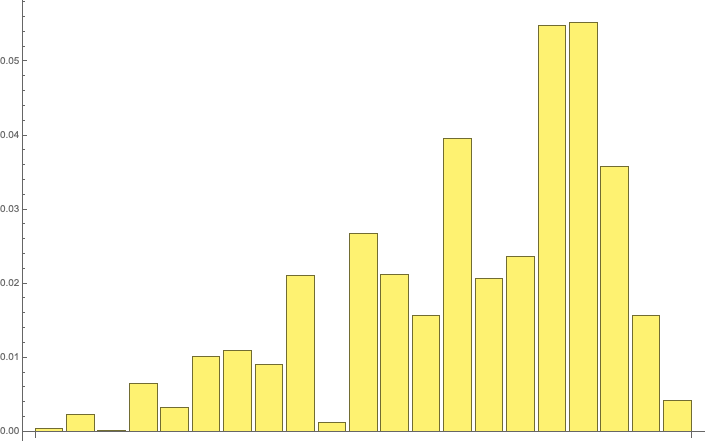}
    \caption{Histograms showing the exact and absolute values of Eq.(\ref{plotclebsch}) for $k=33$
    and $(p_2,q_2)$ values $(95,27), (93,28),\ldots (55,47)$  }
    \label{fig:2dhistogramsk33}
\end{figure}

Finally, Fig.\ref{fig:2dhistogramsk33}, we gives the actual (signed) reduced CG on the left, and the absolute value of this CG on the right, for the 
reduced CG of Eq.(\ref{plotclebsch}) and $(p_2,q_2)$ irreps
with $k=33$.  The reduced CGs are given in left to right order for the $(p_2,q_2)$ values $(95,27), (93,28),\ldots (55,47)$.

From a computational perspective, 
Mathematica\texttrademark\ has a built in function to evaluate $6j$ coefficients, and so can be used 
very effectively to evaluate matrix elements and the $9j$ symbol as a sum of $6j$ symbols.
The code runs very smoothly, with calculations of CGs taking as little as a fraction of a second.
For large values of $k$, the code can take upwards of 90 seconds but calculation time 
depends on the specific combination of $k$ value, quantum numbers and angular momentum values used for the reduced CG.

\newpage

\bigskip

\bibliographystyle{unsrt}
\bibliography{testbib}

\begin{thebibliography}{10}

\bibitem{varshalovich1988quantum}
Dmitri{\u\i}~Aleksandrovich Varshalovich, Anatolij~Nikolaevi{\v{c}} Moskalev,
  and Valerii~Kel'manovich Khersonskii.
\newblock {\em Quantum theory of angular momentum}.
\newblock World Scientific, 1988.

\bibitem{rose1995elementary}
Morris~Edgar Rose.
\newblock {\em Elementary theory of angular momentum}.
\newblock Courier Corporation, 1995.

\bibitem{edmonds2016angular}
Alan~Robert Edmonds.
\newblock {\em Angular momentum in quantum mechanics}.
\newblock Princeton university press, 2016.

\bibitem{king2002multiplicity}
RC~King and BG~Wybourne.
\newblock Multiplicity-free tensor products of irreducible representations of
  the exceptional lie groups.
\newblock {\em Journal of Physics A: Mathematical and General}, 35(15):3489,
  2002.

\bibitem{van1997coupling}
J~Van~der Jeugt.
\newblock Coupling coefficients for lie algebra representations and addition
  formulas for special functions.
\newblock {\em Journal of Mathematical Physics}, 38(5):2728--2740, 1997.

\bibitem{caprio2009construction}
MA~Caprio, DJ~Rowe, and Trevor~A Welsh.
\newblock Construction of $so (5)\supset so (3)$ spherical harmonics and
  clebsch--gordan coefficients.
\newblock {\em Computer Physics Communications}, 180(7):1150--1163, 2009.

\bibitem{rowe2004spherical}
DJ~Rowe, PS~Turner, and J~Repka.
\newblock Spherical harmonics and basic coupling coefficients for the group so
  (5) in an so (3) basis.
\newblock {\em Journal of mathematical physics}, 45(7):2761--2784, 2004.

\bibitem{hecht1983some}
KT~Hecht and Y~Suzuki.
\newblock Some special $su (3) \supset r (3)$ wigner coefficients and their
  application.
\newblock {\em Journal of Mathematical Physics}, 24(4):785--792, 1983.

\bibitem{chen2002group}
Jin-Quan Chen, Jialun Ping, and Fan Wang.
\newblock {\em Group representation theory for physicists}.
\newblock World Scientific Publishing Company, 2002.

\bibitem{weichselbaum2012non}
Andreas Weichselbaum.
\newblock Non-abelian symmetries in tensor networks: A quantum symmetry space
  approach.
\newblock {\em Annals of Physics}, 327(12):2972--3047, 2012.

\bibitem{everdingcalculation}
Lukas Everding.
\newblock Calculation of clebsch-gordan coefficients via weyl group symmetry,
  hbsc. thesis, munich, 2011.

\bibitem{derome1966symmetry}
Jean-Robert Derome.
\newblock Symmetry properties of the 3 j-symbols for an arbitrary group.
\newblock {\em Journal of Mathematical Physics}, 7(4):612--615, 1966.

\bibitem{smorodinskii1972clebsch}
Ya~A Smorodinskii and Leonid~A Shelepin.
\newblock {\em Soviet Physics Uspekhi}, 15(1):1, 1972.

\bibitem{smorodinskii1978clebsch}
Ya~A Smorodinskii.
\newblock {\em Soviet Journal of Experimental and Theoretical Physics}, 48:403,
  1978.

\bibitem{regge1958symmetry}
Tullio Regge.
\newblock Symmetry properties of clebsch-gordan coefficients.
\newblock {\em Il Nuovo Cimento (1955-1965)}, 10(3):544--545, 1958.

\bibitem{draayer1973wigner}
JP~Draayer and Yoshimi Akiyama.
\newblock {\em Journal of Mathematical Physics}, 14(12):1904--1912, 1973.

\bibitem{alex2011numerical}
Arne Alex, Matthias Kalus, Alan Huckleberry, and Jan von Delft.
\newblock {\em Journal of Mathematical Physics}, 52(2):023507, 2011.

\bibitem{bahri2004programs}
C~Bahri, DJ~Rowe, and JP~Draayer.
\newblock Programs for generating clebsch--gordan coefficients of su (3) in su
  (2) and so (3) bases.
\newblock {\em Computer physics communications}, 159(2):121--143, 2004.

\bibitem{stancu20163}
Fl~Stancu.
\newblock {\em Nuclear Physics A}, 945:144--152, 2016.

\bibitem{de1995asymptotic}
Hubert de~Guise and David~J Rowe.
\newblock Asymptotic clebsch--gordan coefficients.
\newblock {\em Journal of Mathematical Physics}, 36(12):6991--7008, 1995.

\bibitem{de1998clebsch}
Hubert de~Guise and David~J Rowe.
\newblock Clebsch--gordan coefficients in the asymptotic limit.
\newblock {\em Journal of Mathematical Physics}, 39(2):1087--1106, 1998.

\bibitem{hecht1965su3}
KT~Hecht.
\newblock {\em Nuclear Physics}, 22:1--36, 1965.

\bibitem{rowe1999representations}
DJ~Rowe, BC~Sanders, and H~de~Guise.
\newblock {\em Journal of Mathematical Physics}, 40:3604, 1999.

\bibitem{schwinger2015angular}
Julian Schwinger.
\newblock {\em On Angular Momentum}.
\newblock Courier Dover Publications, 2015.

\bibitem{rowe1997algebraic}
DJ~Rowe and J~Repka.
\newblock {\em Journal of Mathematical Physics}, 38(8):4363--4388, 1997.

\bibitem{williams1996su3}
H~Thomas Williams.
\newblock Su3 isoscalar factors.
\newblock {\em Journal of Mathematical Physics}, 37(8):4187--4198, 1996.

\bibitem{de1963octet}
JJ~De~Swart.
\newblock The octet model and its clebsch-gordan coefficients.
\newblock {\em Reviews of Modern Physics}, 35(4):916, 1963.

\bibitem{de2018simple}
Hubert de~Guise, Olivia Di~Matteo, and Luis~L S{\'a}nchez-Soto.
\newblock {\em Physical Review A}, 97(2):022328, 2018.

\bibitem{russell2017direct}
Nicholas~J Russell, Levon Chakhmakhchyan, Jeremy~L O?Brien, and Anthony Laing.
\newblock Direct dialling of haar random unitary matrices.
\newblock {\em New journal of physics}, 19(3):033007, 2017.

\bibitem{murnaghan1962unitary}
Francis~Dominic Murnaghan.
\newblock {\em The unitary and rotation groups}, volume~3.
\newblock Spartan books, 1962.

\bibitem{speiser1964theory}
D~Speiser.
\newblock {\em Theory of compact Lie groups and some applications to elementary
  particle physics}.
\newblock Gordon and Breach, 1964.

\bibitem{wesslen2008geometric}
Maria~SM Wessl{\'e}n.
\newblock A geometric description of tensor product decompositions in su (3).
\newblock {\em Journal of Mathematical Physics}, 49(7):073506, 2008.

\bibitem{o1982closed}
Michael~F O'Reilly.
\newblock A closed formula for the product of irreducible representations of su
  (3).
\newblock {\em Journal of Mathematical Physics}, 23(11):2022--2028, 1982.

\bibitem{campoamor2019external}
R~Campoamor-Stursberg.
\newblock The external labelling problem and clebsch--gordan series of
  semisimple lie algebras.
\newblock {\em Journal of Physics A: Mathematical and Theoretical},
  52(12):125201, 2019.

\bibitem{campoamor2019some}
R~Campoamor-Stursberg.
\newblock Some empirical formulae for the degeneracy separation in the
  clebsch-gordan problem of.
\newblock In {\em Journal of Physics: Conference Series}, volume 1194, page
  012019. IOP Publishing, 2019.

\bibitem{martins2019correspondence}
Alex Clesio~Nunes Martins, Andrei~B Klimov, and Hubert de~Guise.
\newblock Correspondence rules for wigner functions over su (3)/u (2).
\newblock {\em Journal of Physics A: Mathematical and Theoretical}, 2019.

\bibitem{klimov2017generalized}
Andrei~B Klimov, Jos{\'e}~Luis Romero, and Hubert De~Guise.
\newblock Generalized su (2) covariant wigner functions and some of their
  applications.
\newblock {\em Journal of Physics A: Mathematical and Theoretical},
  50(32):323001, 2017.

\bibitem{su3githubcode}
Mark~W Suffak, Alex~Cl\'{e}sio Nunes-Martins, and H~de~Guise.
\newblock https://github.com/marksuffak/SU-3-CG-Code, 2018.

\bibitem{sprung2009asymptotic}
DWL Sprung, W~van Dijk, J~Martorell, and DB~Criger.
\newblock Asymptotic approximations to clebsch-gordan coefficients from a
  tight-binding model.
\newblock {\em American Journal of Physics}, 77(6):552--561, 2009.

\bibitem{wigner2012group}
Eugene Wigner.
\newblock {\em Group theory: and its application to the quantum mechanics of
  atomic spectra}, volume~5.
\newblock Elsevier, 2012.

\bibitem{brussaard1957classical}
PJ~Brussaard and H~A\_ Tolhoek.
\newblock Classical limits of clebsch-gordan coefficients, racah coefficients
  and $d_{mn}^\ell (\varphi, \vartheta, \psi)-functions$.
\newblock {\em Physica}, 23(6-10):955--971, 1957.

\bibitem{ponzano1968semiclassical}
G~Ponzano and T~Regge.
\newblock Semiclassical limit of racah coefficients in spectroscopic and group
  theoretical methods in physics (bloch ed.), 1968.

\bibitem{reinsch1999asymptotics}
Matthias~W Reinsch and James~J Morehead.
\newblock Asymptotics of clebsch--gordan coefficients.
\newblock {\em Journal of Mathematical Physics}, 40(10):4782--4806, 1999.

\bibitem{rowe2010shifted}
David~J Rowe and Hubert De~Guise.
\newblock The shifted harmonic approximation and asymptotic su (2) and su (1,
  1) clebsch--gordan coefficients.
\newblock {\em Journal of Physics A: Mathematical and Theoretical},
  43(50):505307, 2010.

\end{thebibliography}
\end{document}